%% file: SMP-18-001_temp.tex
\pdfoutput=1

\documentclass[11pt,twoside,a4paper,cmspaper,final,collab]{cms-tdr}
\def\svnVersion{2702406-D}\def\svnDate{2019/06/04}\def\cmsCernNoTag{CERN-EP-2018-333}\def\cmsCernDate{\today}\def\cmsMessage{"Published in Physics Letters B as \href{http://dx.doi.org/10.1016/j.physletb.2019.05.042}{\doi{10.1016/j.physletb.2019.05.042}.}"}

\begin{document}\cmsNoteHeader{SMP-18-001}

\hyphenation{had-ron-i-za-tion}
\hyphenation{cal-or-i-me-ter}
\hyphenation{de-vices}
\RCS$HeadURL$
\RCS$Id$
\newlength\cmsFigWidth
\ifthenelse{\boolean{cms@external}}{\setlength\cmsFigWidth{0.49\textwidth}}{\setlength\cmsFigWidth{0.65\textwidth}}
\newlength\cmsFigWidthSplit
\ifthenelse{\boolean{cms@external}}{\setlength\cmsFigWidthSplit{0.49\textwidth}}{\setlength\cmsFigWidthSplit{0.49\textwidth}}
\ifthenelse{\boolean{cms@external}}{\providecommand{\cmsLeft}{upper\xspace}}{\providecommand{\cmsLeft}{left\xspace}}
\ifthenelse{\boolean{cms@external}}{\providecommand{\cmsRight}{lower\xspace}}{\providecommand{\cmsRight}{right\xspace}}
\ifthenelse{\boolean{cms@external}}{\providecommand{\cmsTable}[1]{\resizebox{\columnwidth}{!}{#1}}}{\providecommand{\cmsTable}[1]{#1}}
\ifthenelse{\boolean{cms@external}}{\providecommand{\cmsULeft}{upper left\xspace}}{\providecommand{\cmsULeft}{left\xspace}}
\ifthenelse{\boolean{cms@external}}{\providecommand{\cmsURight}{upper right\xspace}}{\providecommand{\cmsURight}{second from left\xspace}}
\ifthenelse{\boolean{cms@external}}{\providecommand{\cmsLLeft}{lower left\xspace}}{\providecommand{\cmsLLeft}{third from left\xspace}}
\ifthenelse{\boolean{cms@external}}{\providecommand{\cmsLRight}{lower right\xspace}}{\providecommand{\cmsLRight}{right\xspace}}

\newcommand{\ZZ}{\ensuremath{\PZ\PZ}\xspace}
\newcommand{\Zg}{\ensuremath{\PZ\PGg}\xspace}
\newcommand{\WZ}{\ensuremath{\PW\PZ}\xspace}
\newcommand{\WZjj}{\ensuremath{\PW\PZ\mathrm{jj}}\xspace}
\newcommand{\EW}{\ensuremath{\text{EW}}\xspace}
\newcommand{\EWWZ}{\ensuremath{\EW\ \PW\PZ}\xspace}
\newcommand{\QCDWZ}{\ensuremath{\mathrm{QCD}\ \PW\PZ}\xspace}
\newcommand{\pp}{\ensuremath{\Pp\Pp}\xspace}
\hyphenation{ATGCs}
\newcommand{\eee}{\ensuremath{\Pe\Pe\Pe}\xspace}
\newcommand{\eem}{\ensuremath{\Pe\Pe\PGm}\xspace}
\newcommand{\emm}{\ensuremath{\PGm\PGm\Pe}\xspace}
\newcommand{\mmm}{\ensuremath{\PGm\PGm\PGm}\xspace}
\newcommand{\Zpj}{\ensuremath{{\PZ}\mathrm{+jet}}\xspace}
\newcommand{\tZq}{\ensuremath{\PQt\PZ\PQq}\xspace}
\newcommand{\cPV}{\HepParticle{V}{}{}\xspace}
\newcommand{\VVV}{\ensuremath{\cPV\cPV\cPV}\xspace}
\newcommand{\jet}{\ensuremath{\mathrm{j}}}
\newcommand{\detajj}{\ensuremath{\Delta\eta(\mathrm{j}_{1}, \mathrm{j}_{2})}}
\newcommand{\etajj}{\ensuremath{\Delta\eta_{\mathrm{jj}}}}
\newcommand{\mjj}{\ensuremath{m_{\mathrm{jj}}}}
\newcommand{\zepl}{\ensuremath{\eta^{3\ell} - (\eta^{\jet_{1}} + \eta^{\jet_{2}})/2}}
\newcommand{\etas}{\ensuremath{\eta^{*}_{3\ell}}}
\newcommand{\mt}{\ensuremath{m_{\mathrm{T}}(\PW\PZ\xspace)}}
\newcommand{\MG}{\MGvATNLO}
\newcommand{\VBFNLO}{\textsc{Vbfnlo}\xspace}
\newcommand{\MocaPlus}{\textsc{MoCaNLO}+\textsc{Recola}\xspace}
\newcommand{\Rivet}{\textsc{Rivet}\xspace}
\newcommand{\SUtwo}{SU(2)}
\newcommand{\Uone}{U(1)}
\newcommand{\FxFx}{{\textsc{FxFx}}\xspace}
\newcommand{\muR}{\ensuremath{\mu_{\text{R}}}\xspace}
\newcommand{\muF}{\ensuremath{\mu_{\text{F}}}\xspace}
\newcommand{\scale}{\ensuremath{\,\text{(scale)}}}
\newcommand{\PDF}{\ensuremath{\,\text{(PDF)}}}
\newcommand{\Hpjj}{\ensuremath{{\Hpm}\mathrm{+jj}}\xspace}

\cmsNoteHeader{SMP-18-001}
\title{Measurement of electroweak \WZ boson production and search for new physics in \WZ $+$ two jets events in {\pp} collisions at $\sqrt{s} = 13\TeV$}

\date{\today}
\abstract{A measurement of \WZ electroweak (EW) vector boson scattering is presented.
The measurement is performed in the leptonic decay modes $\WZ \to \ell\nu\ell'\ell'$,
where $\ell, \ell' = \Pe$, $\mu$.
The analysis is based on a data sample of proton-proton
collisions at $\sqrt{s} = 13\TeV$ at the LHC collected with the CMS detector and corresponding to an integrated luminosity of $35.9\fbinv$.
The \WZ plus two jet production cross section
is measured in fiducial regions with enhanced contributions from
EW production and found to be consistent with standard model predictions.
The \EWWZ production in association with two jets is measured with an observed (expected) significance of 2.2 (2.5) standard deviations.
Constraints on charged Higgs boson production and on anomalous quartic gauge couplings in terms of
dimension-eight effective field theory operators are
also presented.
}

\hypersetup{
pdfauthor={CMS Collaboration},
pdftitle={Measurement of electroweak WZ boson production and search for new physics in WZ + two jets events in pp collisions at sqrt(s) = 13 TeV},
pdfsubject={CMS},
pdfkeywords={CMS, physics, SM, WZ, VBS}}

\maketitle

\section{Introduction}
\label{sec:introduction}
The discovery of a scalar boson with couplings consistent with those of the standard model (SM)
Higgs boson ({\PH}) by the ATLAS and CMS Collaborations~\cite{Aad:2012tfa,Chatrchyan:2012xdj,Chatrchyan:2013lba} at the CERN LHC
provides evidence that the {\PW} and {\cPZ} bosons acquire mass through the
Brout-Englert-Higgs mechanism~\cite{PhysRevLett.13.321,Higgs:1964ia,PhysRevLett.13.508,PhysRevLett.13.585,PhysRev.145.1156,PhysRev.155.1554}.
However, current measurements of the Higgs boson
couplings~\cite{Khachatryan:2016vau,Sirunyan:2018koj}
do not preclude the existence of scalar
isospin doublets, triplets, or higher isospin representations
alongside the single isospin doublet field
responsible for breaking the electroweak (EW) symmetry in the SM~\cite{Chiang:2018cgb,Chowdhury:2017aav}.
In addition to their couplings to the Higgs boson,
the non-Abelian nature of the EW sector of the SM leads to
quartic and triple self-interactions of the massive vector bosons.
Physics beyond the SM in the EW sector
is expected to include
interactions with the vector and Higgs bosons that modify their effective couplings.
Characterizing the self-interactions of the
vector bosons is thus of great importance.

\begin{figure}[htbp]
  \centering
  {\includegraphics[width=0.21\textwidth]{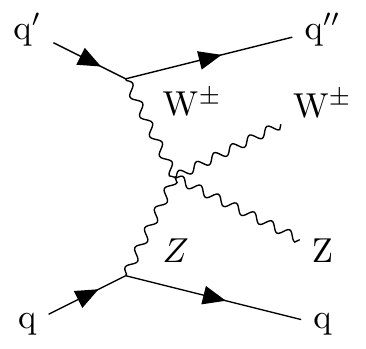}}\qquad
   {\includegraphics[width=0.21\textwidth]{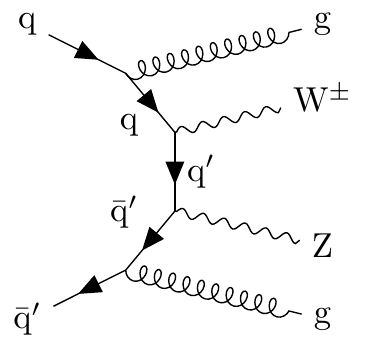}}\qquad
   {\includegraphics[width=0.21\textwidth]{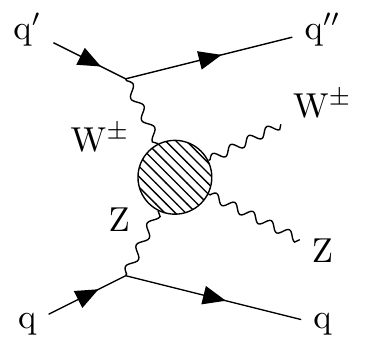}}\qquad
   {\includegraphics[width=0.21\textwidth]{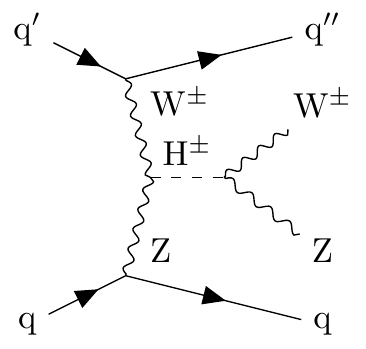}}\qquad
  \caption{Representative Feynman diagrams for \WZjj production in the SM and beyond the SM.
  The EW-induced component of \WZ production includes quartic interactions (\cmsULeft) of the vector bosons.
  This is distinguishable from QCD-induced production (\cmsURight) through kinematic variables.
  New physics in the EW sector modifying the quartic coupling
  can be parameterized in terms of dimension-eight effective field theory operators (\cmsLLeft).
  Specific models modifying this interaction include those predicting charged Higgs bosons (\cmsLRight).
  }
 \label{fig:feynmanDiagrams}
\end{figure}

The total \WZ production cross section in proton-proton (\pp) collisions
has been measured in the leptonic decay modes by the ATLAS and CMS Collaborations
at 7, 8, and
13\TeV~\cite{Aad:2012twa,Aad:2016ett,Khachatryan:2016tgp,Khachatryan:2016poo,Aaboud:2019gxl},
and limits on anomalous triple gauge couplings~\cite{Hagiwara:1989mx} are presented in
Refs.~\cite{Aad:2016ett,Khachatryan:2016poo,Sirunyan:2019bez}.
Constraints on anomalous quartic gauge couplings (aQGC)~\cite{Eboli:2006wa}
are presented by the ATLAS Collaboration at 8\TeV in Ref.~\cite{Aad:2016ett}.
At the LHC, quartic \WZ interactions are accessible through triple vector boson production or via vector boson scattering (VBS),
where vector bosons are radiated from the incoming quarks before interacting,
as illustrated in Fig.~\ref{fig:feynmanDiagrams}~(\cmsULeft).
The VBS processes form a distinct experimental signature characterized by
the {\PW} and {\cPZ} bosons with two forward,
high-momentum jets, arising from the hadronization of
two quarks.
They are part of an important subclass of processes contributing to
\WZ plus two jet (\WZjj) production that proceeds via the EW interaction at tree level,
$\mathcal{O}(\alpha^4)$, referred to as EW-induced \WZjj production, or simply \EWWZ production.
An additional contribution to the \WZjj state proceeds via quantum chromodynamics
(QCD) radiation of partons from
an incoming quark or gluon, shown in Fig.~\ref{fig:feynmanDiagrams}~(\cmsURight),
leading to tree-level contributions at $\mathcal{O}(\alpha^2\alpS^2)$.
This class of processes is referred to as QCD-induced \WZjj production (or \QCDWZ).

The first study of \EWWZ production at the LHC was performed by the ATLAS Collaboration
at 8\TeV~\cite{Aad:2016ett}. A measurement at 13\TeV
with an observed statistical significance for the \EWWZ process greater than 5 standard deviations has recently been reported and submitted for publication by the ATLAS Collaboration~\cite{Aaboud:2018ddq}.  This letter reports searches for \EWWZ production in the SM and for new physics modifying the $\PW\PW\PZ\PZ$ coupling in \pp collisions at $\sqrt{s} = 13\TeV$.  Two fiducial \WZjj cross sections are presented, both in phase spaces with enhanced contributions from the \EWWZ process.  The data sample corresponds to an integrated luminosity of 35.9\fbinv collected with the CMS detector~\cite{Chatrchyan:2008zzk} at the CERN LHC in 2016.  The analysis selects events with exactly three leptons (electrons or muons),
missing transverse momentum {\ptmiss},
and two jets at high pseudorapidity $\eta$ with a large dijet system invariant mass {\mjj},
characteristic of VBS processes.
The kinematic variables of the two forward and high momentum jets,
including $\eta$ separation and {\mjj}, are used
to identify the \EWWZ component of \WZjj production.
An excess of events with respect to the SM prediction could indicate contributions from
additional gauge boson or vector resonances~\cite{Delgado:2017cls},
charged scalar or Higgs bosons~\cite{Kilian:2015opv},
or it could suggest that the gauge or Higgs bosons are not elementary~\cite{Csaki:2015hcd}.
We study such deviations in terms
of aQGCs in the generalized framework of
dimension-eight effective field theory operators, Fig.~\ref{fig:feynmanDiagrams} (\cmsLLeft),
and in terms of charged Higgs bosons, Fig.~\ref{fig:feynmanDiagrams} (\cmsLRight),
and we place limits on their production cross sections and operator couplings.

\section{The CMS detector}
\label{sec:cms}

The central feature of the CMS apparatus is a superconducting solenoid of 6\unit{m} internal diameter, providing a magnetic field of 3.8\unit{T}. Within the solenoid volume are silicon pixel and strip tracking detectors, a lead tungstate crystal electromagnetic calorimeter (ECAL), and a brass and scintillator hadron calorimeter (HCAL), each composed of a barrel and two endcap sections. Forward calorimeters extend the $\eta$ coverage provided by the barrel and endcap detectors up to $\abs{\eta} < 5$. Muons are measured in gas-ionization detectors embedded in the steel flux-return yoke outside the solenoid.

Events of interest are selected using a two-level trigger system~\cite{Khachatryan:2016bia}. The first level of the CMS trigger system, composed of custom hardware processors, uses information from the calorimeters and muon detectors to select events of interest in a fixed time interval of 3.2\mus. The high-level trigger processor farm further decreases the event rate from around 100\unit{kHz} to less than 1\unit{kHz}, before data storage~\cite{Khachatryan:2016bia}.

A more detailed description of the CMS detector, together with a definition of the coordinate system used and the relevant kinematic variables, can be found in Ref.~\cite{Chatrchyan:2008zzk}.

\section{Signal and background simulation}
\label{sec:mc}

Several Monte Carlo (MC) event generators are used to simulate the signal and
background processes.

The EW-induced production of \WZ boson pairs and two final-state quarks, Fig.~\ref{fig:feynmanDiagrams} (\cmsULeft),
where the {\PW} and {\cPZ} bosons decay leptonically,
is simulated at leading order (LO) in perturbative QCD using
\MG~v2.4.2~\cite{MGatNLO}.
The MC simulation includes all contributions to the three-lepton final state at
$\mathcal{O}(\alpha^6)$, with the condition that the mass of {\PW} boson
be within $30\GeV$ of its on-shell value from Ref.~\cite{Tanabashi:2018oca}.
The resonant {\PW} boson is decayed using \textsc{MadSpin}~\cite{Artoisenet:2012st}.
Triboson processes, where the \WZ boson pair is accompanied by
a third vector boson that decays into jets, are included in the MC simulation,
but account for well below 1\% of the event yield for the selections described
in Section~\ref{sec:eventselection}.
Contributions with an initial-state {\cPqb} quark are excluded from
this MC simulation since
they are considered part of the {\tZq} background process.
The predictions from \MG are cross-checked with LO predictions from the
event generators \VBFNLO~3.0~\cite{VBFNLO} and
\SHERPA~v2.2.4~\cite{Gleisberg:2008ta,Gleisberg:2003xi},
and with fixed-order calculations from
\MocaPlus~\cite{leshouches2017,Recola}. Agreement is obtained when using equivalent configurations
of input parameters, including couplings, particle masses and widths, and the choice of
renormalization ($\muR$) and
factorization scales ($\muF$).

Several MC simulations of the \QCDWZ process, Fig.~\ref{fig:feynmanDiagrams} (\cmsURight), are considered.
The simulations are inclusive in the number of jets associated with the
leptonically decaying {\PW} and {\cPZ} bosons, and therefore
comprise the full \WZjj state.
The primary MC simulation is simulated at
LO with \MG~v2.4.2, with contributions to \WZ production with up to three outgoing partons
included in the matrix element calculation.
The different jet multiplicities are merged using the MLM scheme~\cite{MLMmerging}.
A next-to-leading order (NLO) MC simulation from \MG~v2.3.3
with zero or one outgoing partons at Born level, merged using the \FxFx scheme~\cite{Frederix:2012ps},
and an inclusive NLO simulation from \POWHEG2.0~\cite{Melia:2011tj,Nason:2004rx,Frixione:2007vw,powheg:2010}
are also utilized.
The LO MC simulation with MLM merging, referred to as the MLM-merged simulation,
is used as the central prediction for the analysis because of its inclusion of
\WZ plus three-parton contributions at tree level, which are relevant
to \WZjj production.
The other MC simulations,
used to assess the modeling uncertainty in the \QCDWZ process,
are referred to as the \FxFx-merged
and the \POWHEG simulations, respectively.
Each MC simulation is normalized to the NLO cross section from \POWHEG2.0.

In addition to the \EWWZ and \QCDWZ processes, which at tree level are
$\mathcal{O}(\alpha^4)$ and $\mathcal{O}(\alpha^2\alpS^2)$ respectively,
a smaller contribution at $\mathcal{O}(\alpha^3\alpS)$
contributes to the \WZjj state. We refer to this contribution as the
interference term. It is evaluated using MC simulations of particle-level
events generated with \MG~v2.6.0. The process is simulated with the dynamic $\muR$
and $\muF$ set to the maximum outgoing quark \pt per event, and with fixed
scales $\muR = \muF = m_{\PW}$, where $m_{\PW}$ is the world average value of the
$\PW$ boson mass, taken from Ref.~\cite{Tanabashi:2018oca}.

The associated production of a $\cPZ$ boson and a single top quark, referred to as {\tZq} production,
is simulated at NLO in the four-flavor scheme using \MG~v2.3.3.
The MC simulation is normalized using a cross section computed at NLO with \MG in the five-flavor scheme,
following the procedure of Ref.~\cite{Sirunyan:2017nbr}.
The production of $\cPZ$ boson pairs via $\Pq\Paq$ annihilation is generated at NLO in perturbative QCD with
\POWHEG2.0 while the $\Pg\Pg \to \ZZ$ process is simulated at LO with \MCFM7.0~\cite{Campbell:2011bn}.
The $\ZZ$ simulations are normalized to the cross section calculated at next-to-next-to-leading order
for $\Pq\Paq \to \ZZ$ with MATRIX~\cite{Cascioli:2014yka,Grazzini:2017mhc} ($K$ factor 1.1)
and at NLO for $\Pg\Pg \to \ZZ$ \cite{Caola:2015psa} ($K$ factor 1.7).
The \EW production of \Z boson pairs and two final-state quarks,
where the \Z bosons decay leptonically, is simulated at LO using \MG~v2.3.3.
Background from $\Zg$, $\ttbar\text{V}$ ($\ttbar\PW$, $\ttbar\cPZ$),
and triboson events {\VVV} ($\PW\PW\cPZ$, $\PW\cPZ\cPZ$, $\cPZ\cPZ\cPZ$)
are generated at NLO with \MG~v2.3.3, with the vector bosons generated on-shell
and decayed via \textsc{MadSpin}.

The simulation of the aQGC processes is performed at LO using \MG~v2.4.2 and employs matrix element
reweighting to obtain a finely spaced grid of parameters for each of the anomalous couplings
operators probed by the analysis. The configuration of input parameters is equivalent to that used for the
\EWWZ simulation described previously.
The production of charged Higgs bosons in the
Georgi--Machacek (GM) model~\cite{GEORGI1985463}
is simulated at LO using \MG~v2.3.3 and normalized using the next-to-next-to-leading order
cross sections reported in Ref.~\cite{Zaro:2002500}.

The \PYTHIA~v8.212~\cite{Sjostrand:2006za,Sjostrand:2015} package
is used for parton showering, hadronization, and
underlying event simulation, with parameters set by the CUETP8M1
tune~\cite{Khachatryan:2015pea} for all simulated samples.
For the \EWWZ process, comparisons are made at particle-level with the parton shower
and hadronization
of {\SHERPA} and with \HERWIG~v7.1~\cite{Bellm:2015jjp,Bahr:2008pv}.
For all MC simulations used in this analysis, the NNPDF3.0~\cite{NNPDF2015} set of
parton distribution functions (PDFs) is used, with PDFs calculated to the same
order in perturbative QCD as the hard scattering process.

The detector response is simulated using a detailed
description of the CMS detector implemented in the \GEANTfour
package~\cite{GEANT, Geant2}. The simulated events are  reconstructed
using the same algorithms used for the data.
The simulated samples include additional interactions in the same and neighboring bunch crossings,
referred to as pileup.
Simulated events are weighted so the pileup distribution reproduces that observed in
the data, which has an average of about 23 interactions per bunch
crossing.

\section{Event reconstruction}
\label{sec:eventreconstruction}
In this analysis, the particle-flow (PF) event reconstruction algorithm~\cite{CMS-PRF-14-001} is used.
The PF algorithm aims to reconstruct and identify each individual particle as a physics object in an event, with an optimized combination of information from the various elements of the CMS detector. The energy of photons is obtained from the ECAL measurement. The energy of electrons is determined from a combination of the electron momentum at the primary interaction vertex as determined by the tracker, the energy of the corresponding ECAL cluster, and the energy sum of all bremsstrahlung photons spatially compatible with originating from the electron track. The energy of muons is obtained from the curvature of the corresponding track. The energy of charged hadrons is determined from a combination of their momentum measured in the tracker and the matching ECAL and HCAL energy deposits, corrected for zero-suppression effects and for the response function of the calorimeters to hadronic showers. Finally, the energy of neutral hadrons is obtained from the corresponding corrected ECAL and HCAL energies.

The reconstructed vertex with the largest value of summed physics-object $\pt^2$
(where \pt is the transverse momentum) is the primary $\Pp\Pp$ interaction vertex. The physics objects are the jets, clustered using a jet finding algorithm~\cite{Cacciari:2008gp,Cacciari:2011ma} with the tracks assigned to the vertex as inputs, and the associated \ptmiss, taken as the negative vector sum of the $\pt^{\jet}$ of those jets.

Electrons are reconstructed within the geometrical acceptance $\abs{\eta^{\Pe}} < 2.5$.
The reconstruction combines the information from clusters of energy deposits in the ECAL and the
trajectory in the tracker~\cite{Khachatryan:2015hwa}.
To reduce the electron misidentification rate, electron candidates
are subjected to additional identification criteria based on the distribution
of the electromagnetic shower in the ECAL, the relative amount of energy deposited in the
HCAL, a matching of the trajectory of an electron track with the cluster in the
ECAL, and its consistency with originating from the selected primary vertex. Candidates that
are identified as originating from photon conversions in the detector material are removed.

Muons are reconstructed within $\abs{\eta^{\mu}} < 2.4$~\cite{Sirunyan:2018fpa}.
The reconstruction combines the information from both the tracker and the
muon spectrometer.
The muons are selected from among the reconstructed muon track candidates
by applying minimal quality requirements on the track components in the muon system
and by ensuring that muons are associated with small energy deposits in the
calorimeters.

For each lepton track,
the distance of closest approach to the primary vertex in the transverse
plane is required to be less than 0.05\,(0.10)\unit{cm} for electrons in the barrel\,(endcap) region and
0.02\unit{cm} for muons.
The distance along the beamline must be less than 0.1\,(0.2)\unit{cm} for electrons in the barrel\,(endcap) and
0.1\unit{cm} for muons.

Jets are reconstructed using PF objects.
The anti-\kt jet clustering
algorithm~\cite{Cacciari:2008gp} with a distance parameter $R=0.4$ is used.
To exclude electrons and muons from the jet sample, the
jets are required to be separated from the identified leptons by
$\Delta R = \sqrt{\smash[b]{(\Delta \eta)^{2} + (\Delta \phi)^{2}}} > 0.4$,
where $\phi$ is the azimuthal angle in radians.
The CMS standard method for jet energy
corrections~\cite{cmsJEC}
is applied.
These include corrections to the pileup contribution that keep the jet energy correction and the corresponding uncertainty
almost independent of the number of pileup interactions.
In order to reject
jets coming from pileup collisions (pileup jets), a multivariate-based jet identification
algorithm~\cite{CMS-PAS-JME-13-005} is applied.
This algorithm takes advantage of differences in the shape of energy
deposits in a jet cone between jets from hard-scattering and from pileup interactions. The jets are required to have $\pt^{\jet} > 30\GeV$ and
$\abs{\eta^{\jet}}<4.7$.
We identify potential top quark backgrounds by identifying
the {\cPqb} quark produced in its decay via
the combined secondary vertex \cPqb-tagging algorithm with the \textit{tight} working point~\cite{Sirunyan:2017ezt}. The efficiency for selecting {\cPqb} quark jets is $\approx$49\% with
a misidentification probability of $\approx$4\% for {\cPqc} quark jets and $\approx$0.1\% for light-quark and gluon jets.

The isolation of individual electrons or muons is defined relative to their $\pt^{\ell}$
by summing over the \pt of charged hadrons and neutral particles within a cone with radius
$\Delta R < 0.3\,(0.4)$ around the electron (muon) direction
at the interaction vertex:
\begin{linenomath}
\ifthenelse{\boolean{cms@external}}{
\begin{multline*}
I^{\ell} = \Bigg( \sum  \pt^\text{charged}\\
+ \text{max}\Big[ 0, \sum \pt^\text{neutral}
+  \sum \pt^{\gamma} - \pt^{\mathrm{PU}}  \Big] \Bigg) \Big/  \pt^{\ell}.
\end{multline*}
}{
\begin{equation*}
I^{\ell} = \left( \sum  \pt^\text{charged} + \text{max}\left[ 0, \sum \pt^\text{neutral}
                                 +  \sum \pt^{\gamma} - \pt^{\mathrm{PU}}  \right] \right) /  \pt^{\ell}.
\end{equation*}
}
\end{linenomath}
Here, $\sum  \pt^\text{charged}$ is the scalar \pt sum of charged hadrons
originating from the primary vertex.
The $\sum \pt^\text{neutral}$ and $\sum \pt^{\gamma}$ are the
scalar \pt sums for neutral hadrons and photons, respectively.
The neutral contribution to the isolation from pileup events,
$\pt^{\mathrm{PU}}$, is estimated differently for electrons and muons.
For electrons,
$\pt^{\mathrm{PU}} \equiv \rho \, A_\text{eff}$,
where the average transverse momentum
flow density $\rho$ is calculated in each event using the ``jet area'' method~\cite{Cacciari:2007fd}, which defines
$\rho$ as the median of the ratio of the jet transverse momentum to the jet area, $\pt^{\jet}/A_{\jet}$,
for all pileup jets in the event.
The effective area $ A_\text{eff}$ is the geometric area of the isolation cone times an $\eta$-dependent correction factor
that accounts for the residual dependence of the isolation on the pileup.
For muons, $\pt^{\mathrm{PU}} \equiv 0.5 \sum_i \pt^{\mathrm{PU},i}$,
where $i$ runs over the charged hadrons originating from pileup vertices and
the factor 0.5 corrects for the ratio of charged to neutral particle contributions in the isolation cone.
Electrons are considered isolated
if $I^{\Pe} < 0.036,(0.094)$ for the barrel (endcap) region, whereas muons are considered isolated if $I^{\Pgm} < 0.15$, where the values are optimized for
aggressive background rejection while maintaining a reconstruction efficiency of $\approx$70\%.
Relaxed identification criteria are defined
by $I^{\Pgm} < 0.40$ for muons and by relaxed 
track quality and detector-based isolation conditions for electrons.
The overall efficiencies of the reconstruction, identification, and isolation
requirements for the
prompt $\Pe$ or $\Pgm$ are measured in data and simulation
in bins of $\pt^{\ell} $ and $ \abs{\eta^{\ell}} $ using a ``tag--and--probe''
technique~\cite{CMS:2011aa} applied to an inclusive sample of {\cPZ} events.
The data to simulation efficiency ratios are used as scale factors
to correct the simulated event yields.

\section{Event selection}
\label{sec:eventselection}

Collision events are selected by triggers that require the presence of
one or two electrons or muons.
The $\pt^{\ell}$ threshold for the single lepton trigger is 25 (20)\GeV for the electron (muon) trigger.
For the dilepton triggers, with the same or different flavors, the minimum $\pt^{\ell}$ of the leading and subleading leptons are 17\,(17) and 12\,(8)\GeV
for electrons\,(muons), respectively.
The combination of these trigger paths brings the trigger efficiency for selected three-lepton events
to nearly 100\%.
Partial mistiming of signals in the forward region of the electromagnetic calorimeter (ECAL) endcaps
($2.5 < \abs{\eta} < 3.0$) led to early readout for a significant fraction
of events with forward jet activity, and a corresponding
reduction in the level 1 trigger efficiency.
A correction for this effect is determined
in bins of jet $\pt^{\jet}$ and $\eta^{\jet}$
using an unbiased data sample.
This loss of efficiency is about
1\% for {\mjj} of $200\GeV$, increasing to about 15\% for $\mjj > 2\TeV$.

A selected event is required to have three lepton candidates $\ell\ell'\ell'$,
where $\ell, \ell' = \Pe$, $\mu$.
All leptons
must pass the identification and isolation requirements
described in Section~\ref{sec:eventreconstruction}.
The electrons and muons can be directly produced from a {\PW} or {\cPZ} boson decay or from a {\PW} or {\cPZ}
boson with an intermediate $\tau$ lepton decay.
The $\ell'\ell'$ pair consists of two leptons with opposite charge and the same flavor,
as expected for a $\cPZ$ boson candidate.
One of the leptons from the $\cPZ$ boson candidate is required to have $\pt^{\ell'_{1}} > 25\GeV$ and the other
$\pt^{\ell'_{2}} > 15\GeV$.
For events with three same-flavor leptons, two oppositely charged, same-flavor combinations are possible.
The pair with invariant mass closest to $m_{\cPZ} = 91.2\GeV$, the nominal $\cPZ$ boson mass from
Ref.~\cite{Tanabashi:2018oca}, is selected as the $\cPZ$ boson candidate.
The remaining lepton is associated with the $\PW$ boson and must have $\pt^{\ell} > 20\GeV$.
Events containing additional leptons satisfying the relaxed identification criteria with $\pt^{\ell} > 10\GeV$ are rejected.
Because of the neutrino in the final state, the events are required to have $\ptmiss > 30\GeV$.
To reduce contributions from $\ttbar$ events,
the leptons constituting the $\cPZ$ boson candidate are required to have an invariant mass satisfying
$\abs{m_{\ell'\ell'} - m_\cPZ} < 15\GeV $ and events with a
{\cPqb} tagged jet with $\pt^{\cPqb} > 30\,\GeV$ and $\abs{\eta^{\cPqb}} < 2.4$ are vetoed.

The invariant mass of any dilepton
pair $m_{\ell\ell}$ must be greater than 4\GeV.
Such a requirement is necessary in theoretical calculations to avoid divergences
from collinear emission of same-flavor opposite-sign
dilepton pairs, and 4\GeV is chosen to avoid low mass resonances.
The selection is extended to all dilepton pairs to
reduce contributions from backgrounds with soft leptons while having a negligible effect on signal efficiency.
The trilepton invariant mass, $m_{3\ell}$, is required to be more than 100\GeV
to exclude a region where production of $\cPZ$ bosons with final-state photon radiation
is expected to contribute.

\begin{table*}[!ht]
  \centering
  \topcaption{Summary of event selections and fiducial region definitions for the analysis.
    The selections labeled ``EW signal'' and ``Higgs boson'' are applied to data and reconstructed
    simulated events.
    The EW signal selection is used for all measurements except for the charged Higgs boson search that
    uses the selection indicated in the column labeled ``Higgs boson.''
    The \WZjj cross section is reported in the fiducial regions defined by the selections specified in
    the last two columns applied to particle-level simulated events. The variables
    $n_{\jet}$ and $n_{\cPqb}$ refer to the number of
    anti-\kt jets and the number of anti-\kt \cPqb-tagged jets,
    respectively. Other variables are defined in the text.
    }
  \begin{tabular}{ccccc}
  \hline
                                           &  EW signal & Higgs boson & Tight fiducial & Loose fiducial \\
  \hline
  $  \PT^{\ell'_{1}}   $            [GeV]  &  $>$25     & $>$25        & $>$25         & $>$20          \\
  $  \PT^{\ell'_{2}}   $            [GeV]  &  $>$15     & $>$15        & $>$15         & $>$20          \\
  $  \PT^{\ell}     $               [GeV]  &  $>$20     & $>$20        & $>$20         & $>$20          \\
  $\abs{\eta^{\mu}}   $                    &  $<$2.4    & $<$2.4       & $<$2.5        & $<$2.5         \\
  $\abs{\eta^{\Pe}}$                       &  $<$2.5    & $<$2.5       & $<$2.5        & $<$2.5         \\
  $\abs{m_{\ell'\ell'}-m_{\cPZ}}$   [GeV]  &  $<$15     & $<$15        & $<$15         & $<$15          \\
  $m_{3\ell}                $       [GeV]  &  $>$100    & $>$100       & $>$100        & $>$100         \\
  $m_{\ell\ell}           $         [GeV]  &  $>$4      & $>$4         & $>$4          & $>$4           \\
  $\ptmiss                  $       [GeV]  &  $>$30     & $>$30        & \NA           &   \NA          \\
  $\abs{\eta^{\jet}}  $                    &  $<$ 4.7   & $<$4.7       & $<$4.7        & $<$4.7         \\
  $\PT^{\jet}                $      [GeV]  &  $>$50     & $>$30        & $>$50         & $>$30          \\
  $\abs{\Delta R(\jet, \ell)}$             &  $>$0.4    & $>$0.4       & $>$0.4        & $>$0.4         \\
  $n_{\jet}           $                    &  $\ge$2    & $\ge$2       & $\ge$2        & $\ge$2         \\
  $\PT^{\cPqb}          $           [GeV]  &  $>$30     & $>$30        &   \NA         &   \NA          \\
  $\abs{\eta^{\cPqb}}  $                   &  $<$2.4    & $<$2.4       &   \NA         &   \NA          \\
  $n_{\cPqb}       $                       &  $=$0      & $=$0         &   \NA         &   \NA          \\
  $\mjj             $                      &  $>$500    & $>$500       & $>$500        & $>$500         \\
  $\abs{\etajj }$                          &  $>$2.5    & $>$2.5       & $>$2.5        & $>$2.5         \\
  $\abs{ \zepl }$                          &  $<$2.5    & \NA          & $<$2.5       & \NA            \\
  \end{tabular}
  \label{tab:selections}
\end{table*}

Furthermore, the event must have at least two jets with $\pt^{\jet} > 50\GeV$ and $\abs{\eta^{\jet}} < 4.7$.
The jet with the highest $\pt^{\jet}$ is
called the leading jet and the jet with the second-highest $\pt^{\jet}$ the subleading jet.
To exploit the unique signature of the VBS process, these two jets are required to have
$\mjj > 500\GeV$ and $\eta$ separation
$\abs{\detajj} \equiv \abs{\etajj} > 2.5$.
The variable $\etas = \zepl$
of the three-lepton system is additionally required to be between $-2.5$ and 2.5. This selection is
referred to as the ``EW signal selection." The same set of selections, but with no requirement on
{\etas} and with the relaxed requirement $\pt^{j} > 30\GeV$,
is used in searches for charged Higgs bosons and therefore called the ``Higgs boson selection."
A summary of these selections is shown in Table~\ref{tab:selections}.

Sideband regions of events with a similar topology to signal events,
but outside the signal region, are used to constrain the normalization of
the \QCDWZ process in the \EWWZ measurement and in searches for new physics.
We refer to this region as the ``\QCDWZ sideband region.''
It consists of events with $\mjj > 100\GeV$ satisfying all requirements applied to signal events,
but failing at least one of the signal discriminating variables, i.e., $\mjj < 500\GeV$ or
$\abs{\etajj} < 2.5$.
For the \EWWZ measurement, events satisfying $\abs{\etas} > 2.5$ are also selected
in the sideband region.

To reduce the dependence on theoretical predictions,
measurements are reported in two fiducial regions, defined in Table~\ref{tab:selections}.
The ``tight fiducial region'' is defined to be as close as possible to the measurement phase space,
whereas the ``loose fiducial region'' is designed to be easily reproducible
in theoretical calculations or in MC simulations, following the procedure of
Ref.~\cite{leshouches2017}.
The fiducial predictions are defined through selections on particle-level
simulated events using the \Rivet~\cite{Buckley:2010ar} framework, which
provides a toolkit for analyzing simulated events in a model-independent way.
Electrons and muons are required to be prompt (i.e., not from hadron decays),
and those produced in the decay of a $\tau$ lepton
are not considered in the definition
of the fiducial phase space.
The momenta of prompt photons located within a cone of radius $\Delta R = 0.1$ are added to the lepton
momentum to correct for final-state photon radiation, referred to as ``dressing.''
The three highest \pt leptons are selected and associated with the {\PW} and {\cPZ}
bosons with the same procedure used in the data selection.
The fiducial cross section in the \QCDWZ sideband region is defined
following the tight fiducial region of Table~\ref{tab:selections},
with $\mjj > 100\GeV$ and $\mjj < 500\GeV$ or $\abs{\etajj} < 2.5$ or $\abs{\etas} > 2.5$.
Theoretical predictions are evaluated using \MG at LO interfaced to \PYTHIA with the samples
described in Section~\ref{sec:mc}.

\section{Background estimation}
\label{sec:backgrounds}
Background contributions in this analysis are divided into two categories:
background processes with prompt isolated leptons, \eg,
$\cPZ\cPZ$, {\tZq}, $\ttbar\cPZ$;
and background processes with nonprompt leptons from hadrons decaying to leptons inside jets or
jets misidentified as isolated leptons, primarily $\ttbar$ and $\cPZ$+jets.
The background processes with prompt leptons are estimated from MC simulation, whereas backgrounds
with nonprompt leptons from hadronic activity are estimated from data using control samples.
The nonprompt component of the $\Zg$ process,
in which the photon experiences conversion into leptons in the tracker,
is evaluated using MC simulation.

The contribution from \QCDWZ production is estimated with MC simulation.
It is considered signal for the \WZjj cross section measurement,
but is the dominant background for the \EWWZ measurement and in searches for
new physics.
For the \EWWZ measurement
and new physics searches, the normalization of the \QCDWZ process is
constrained by data in the \QCDWZ sideband region.
The cross section predicted by the MLM-merged sample
in the \QCDWZ sideband region is
$18.6 \,^{+2.9}_{-2.3}\scale \pm 1.0\PDF\unit{fb}$,
where the scale and PDF uncertainties are calculated using the procedure
described in Section~\ref{sec:systematics}.
In this region the normalization correction, which is derived from a fit to the data, is
consistent with unity.
The \EWWZ process, considered signal for the \WZjj and \EWWZ
measurements but background to new physics searches, is also
estimated using MC simulation.

The contribution from background processes with nonprompt leptons is
evaluated with data control samples of events satisfying relaxed lepton
identification requirements using the technique described in
Refs.~\cite{Khachatryan:2016tgp,Sirunyan:2017sbn}.
Events satisfying the full analysis selection,
with the exception that one, two, or three leptons pass relaxed identification
requirements but fail the more stringent requirements applied to signal events,
are selected to form relaxed lepton control samples. These control samples are
mutually independent and, additionally, independent from the signal selection.
The small contribution to the relaxed lepton control samples from events with three prompt leptons
is estimated with MC simulation and
subtracted from the event samples.

The expected contribution in the signal region is estimated
using ``loose-to-tight'' efficiency factors
applied to the lepton candidates failing the analysis requirements
in the control region events.
The efficiency factors are calculated from a
sample of $\PZ$+$\ell_{\text{cand}}$ events, where {\cPZ}
denotes a pair of oppositely charged, same-flavor leptons satisfying the
full identification requirements and $\abs{m_{\ell^{+}\ell^{-}} - m_\cPZ} < 10\GeV$,
and $\ell_{\text{cand}}$ is a lepton candidate satisfying the relaxed identification.
The loose-to-tight efficiency factors
are obtained from ratios of events where the $\ell_{\text{cand}}$
object satisfies the full identification requirements
to events where all identification criteria are not satisfied, and
is parameterized as a function of \PT and $\eta$.
A cross-check of the technique is performed by
repeating the procedure with efficiency factors derived from a
sample of events dominated by dijet production. The
loose-to-tight efficiency factors obtained in the two regions
agree to within 30\% for the full \PT and $\eta$ range.

This method is validated in nonoverlapping data samples enriched in Drell--Yan and $\ttbar$ contributions.
The Drell--Yan sample is defined by inverting the selection requirement in $\ptmiss$, and
the $\ttbar$ sample is defined by requiring at least one \cPqb-tagged jet and rejecting events with $\abs{m_{\ell'\ell'} - m_\PZ} < 5\GeV$
while keeping all other requirements for the signal region.
The predictions derived from the relaxed lepton data control samples
agree with the measurements
in the Drell--Yan and $\ttbar$ data samples to within 20\%.

The small size of the loose lepton control samples
and $\Zg$ MC simulation
limit differential predictions in the EW signal region.
Therefore, the combined shape of the estimated nonprompt
and $\Zg$ backgrounds
for both electrons and muons are used as background for the \EWWZ measurement
and in the extraction of constraints on aQGCs.
The normalization of the distributions per channel are taken from the
ratio of the nonprompt ($\Zg$) yield in a single channel to the total nonprompt ($\Zg$) event yield
measured in \WZjj events with no requirements on the dijet system.
These ratios are consistent within the statistical uncertainty with ratios measured
when relaxing the jet \PT requirement in \WZjj events, in \WZ events inclusive in the number of jets,
and in events satisfying the EW signal and \QCDWZ sideband selections.

\section{Systematic uncertainties}
\label{sec:systematics}

The dominant uncertainties in both the cross section measurement
and new physics searches are those associated with
the jet energy scale (JES) and resolution (JER).  The JES and JER
uncertainties are evaluated in simulated events by smearing and
scaling the relevant
observables and propagating the effects to the event selection and
the kinematic variables used in the analysis.
The uncertainty in the event yield in the EW signal selection
due to the JES and JER
is 9\% for \QCDWZ and 5\% for \EWWZ processes.
For the \QCDWZ (\EWWZ) process, the JES uncertainty varies
in the range of 5--25\% (3--15\%) with
increasing values of {\mjj} and $\abs{\etajj}$.

The uncertainties in signal and background processes estimated with
MC simulation are evaluated from the theoretical uncertainties
of the predictions.
Event weights in the MC simulations are used to evaluate
variations of the central prediction.
Scale uncertainties are estimated by independently varying
$\muR$ and $\muF$ by a factor of two from their
nominal values, with the condition that $1/2 \le \muR/\muF \le 2$.
The maximal and minimal variations are obtained
per bin to form a shape-dependent variation band.
The PDF uncertainties are evaluated by combining the predictions per bin from
the fit and $\alpha_s$ variations of the NNPDF3.0 set according to the
procedure described in Ref.~\cite{PDFLHC} for MC replica sets.
The scale and PDF uncertainties are uncorrelated for different signal and
background process and 100\% correlated across bins for the distributions
used to extract results.
For MC simulations normalized to a cross section computed at a higher order
in QCD, the uncertainties are calculated from the order of the MC simulation.

The uncertainty in modeling the \EWWZ and \QCDWZ
processes has a large impact in the \EWWZ measurement.
In addition to the uncertainties from scale and PDF choice,
comparisons of alternative matrix element and parton shower generators are
considered.
The uncertainty in the \QCDWZ process is derived by
comparing the predictions of the MLM-merged simulation and those obtained with the \FxFx-merged simulation,
after fixing the normalization to the observed data
in the \QCDWZ sideband region.
Differences between the predictions of the MC simulations
in the signal region and in the ratio
of the \QCDWZ sideband to the signal region event yields
are considered in the comparisons.
The differences in predictions are generally
within the scale and PDF uncertainties of the MC simulations,
and a 10\% normalization uncertainty is assigned to account for
the observed discrepancies.
The results obtained using the \POWHEG simulation,
which predicts a slightly softer {\mjj} spectrum, are also largely contained
within the theoretical uncertainties considered.
However, because \WZjj events from this simulation arise from
soft radiation from the parton shower, it is
not explicitly considered in the uncertainty evaluation.
For the \EWWZ process, the MC simulations described in Section~\ref{sec:mc}
agree within
the theoretical uncertainties from the PDF and the choice of
$\muR$ and $\muF$
for the kinematic variables considered in the analysis,
so no additional uncertainty is
assigned.

The interference term is evaluated on particle-level simulated events
selected from the MC simulations described in Section~\ref{sec:mc}.
It is positive, and roughly 12\%
of the \EWWZ contribution in the \QCDWZ sideband region and 4\% in the EW signal region
for both MC simulations considered, consistent with the results reported in Ref.~\cite{leshouches2017}.
The ratio of the interference to the \EWWZ
decreases with increasing $\mjj$, consistent with the observations of
Refs.~\cite{leshouches2017,Ballestrero:2018anz}.
These values are used as a symmetric shape uncertainty in the
\EWWZ prediction.
This uncertainty is lower than other theoretical uncertainties
and has a negligible contribution to the uncertainty
in the \EWWZ measurement.

Higher-order \EW corrections in VBS processes are known to be negative and at
the level of tens of percent, with the correction increasing in magnitude
with increasing {\mjj} and $m_{\mathrm{VV}}$~\cite{Biedermann:2016yds}.
We do not apply corrections to the {\WZjj} MC simulation, but we have verified that the
significance of the \EWWZ measurement is insensitive to higher-order \EW corrections by
performing the signal extraction described in Section~\ref{sec:SMWZ} with the {\mjj} predicted
by the \EWWZ MC simulation modified by the corrections from Ref.~\cite{Denner:2019tmn}.
As the relative effect of the EW corrections on SM and anomalous \WZjj production is unknown,
we do not apply corrections to the SM backgrounds or new physics signals for our results.
Because corrections to the SM WZjj production that decrease the expected number of events
at high $m_{\WZ}$ lead to more stringent limits on new physics, this is a conservative approach.

The uncertainties related to the finite number of simulated events, or to the limited
number of events in data control regions, affect the signal and background predictions.
They are uncorrelated
across different samples, and across bins of a single distribution.
The limited number of events in the relaxed lepton control samples used for the
nonprompt background estimate is the dominant contribution to this uncertainty.

The nonprompt background estimate is also affected by systematic uncertainties
from the jet flavor composition of the relaxed lepton control samples
and loose-to-tight extrapolation factors.
The systematic uncertainty in the nonprompt event yield is 30\%
for both electrons and muons, uncorrelated between channels.
It covers the largest difference observed
between the estimated and measured
numbers of events in data control samples enriched in
$\ttbar$ and Drell--Yan contributions and the differences between
using extrapolation factors derived in \Zpj and dijet events.

\begin{table}[htbp]
     \centering
     \topcaption{ The dominant systematic uncertainty contributions in the fiducial
         \WZjj cross section measurement.\label{tab:systematics}}
    \cmsTable{ \begin{tabular}{lc}
 \hline
     Source of syst. uncertainty &      Relative uncertainty in $\sigma_{\WZjj}$ [\%] \\
 \hline
 Jet energy scale                     & $+11\, /-8.1$   \\
 Jet energy resolution                & $+1.9\,/-2.1$   \\
 Prompt background normalization      & $+2.2\,/-2.2$  \\
 Nonprompt normalization              & $+2.5\,/-2.5$  \\
 Nonprompt event count                & $+6.0\,/-5.8$  \\
 Lepton energy scale and eff.         & $+3.5\,/-2.7$  \\
 {\cPqb} tagging                      & $+2.0\,/-1.7$  \\
 Integrated luminosity                & $+3.6\,/-3.0$  \\
 \hline
      \end{tabular}
      }
\end{table}

Systematic uncertainties are less than 1\% for the trigger efficiency and 1--3\% for the
lepton identification and isolation requirements, depending on the lepton flavors.
Other systematic uncertainties are related to the use of simulated samples:
1\% for the effects of pileup and  1--2\% for the \ptmiss reconstruction,
estimated by varying the energies of the PF objects within their uncertainties.
The uncertainty in the {\cPqb} tagging efficiency is 2\% for $\WZ$ events,
which accounts for differences in {\cPqb} tagging efficiencies between MC simulations and data.
The uncertainty in the integrated luminosity of the data sample is 2.5\%~\cite{CMS:2017sdi}.
This uncertainty affects both the signal and the simulated portion of the background estimation,
but does not affect the background estimation from data.

For the extraction of results, log-normal probability density functions are
assumed for the nuisance parameters affecting
the event yields of the various background contributions, whereas systematic uncertainties
that affect the shape of the distributions are represented by nuisance parameters whose
variation results in a continuous perturbation of the spectrum~\cite{Prosper:2011zz} and are assumed
to have a Gaussian probability density function.
A summary of the contribution of each systematic uncertainty to the total
\WZjj cross section measurement is presented in Table~\ref{tab:systematics}.
The impact of each systematic uncertainty in the \WZjj
cross section measurement is obtained by freezing the set of associated nuisance
parameters to their best-fit values and comparing the total uncertainty in the signal strength
to the result from the nominal fit.
The prompt background normalization uncertainty includes the
scale and PDF uncertainties in the background processes estimated using MC simulations.

\section{Fiducial \texorpdfstring{\WZjj}{WZ jj} cross section measurement and search for \texorpdfstring{\EWWZ}{EW WZ} production} \label{sec:SMWZ}
\label{sec:SMWZ}

The cross section for \WZjj production, without separating by production mechanism,
is measured with a combined maximum likelihood fit to the
observed event yields for the EW signal selection.
The likelihood is a combination of individual likelihoods for the four leptonic decay channels
(\eee, \eem, \emm, \mmm) for the
signal and background hypotheses with the statistical and systematic uncertainties in the form
of nuisance parameters.
To minimize the dependence of the result on theoretical predictions,
the likelihood function is built from the event yields per channel
without considering information about the distribution of events in kinematic variables.
The expected event yields for the EW- and QCD-induced \WZjj processes
are taken from the \MG~v2.4.2 predictions.
The \WZjj signal strength $\mu_{\WZjj}$, which is the
ratio of the measured signal yield to the expected number of signal events,
is treated as a free parameter in the fit.

The best-fit value for the \WZjj signal strength is used to obtain a cross section
in the tight fiducial region defined in Table~\ref{tab:selections}.
The measured fiducial \WZjj cross section in this region is
\begin{linenomath}
\begin{equation*}
  \sigma^{\mathrm{fid}}_{\WZjj} =
        3.18^{+0.57}_{-0.52}\stat ^{+0.43}_{-0.36}\syst \unit{fb}
        = 3.18^{+0.71}_{-0.63} \unit{fb}.
\end{equation*}
\end{linenomath}
This result can be compared with the predicted value of
$3.27 \, ^{+0.39}_{-0.32}\scale \pm 0.15\PDF \unit{fb}$.
The \EWWZ and \QCDWZ contributions are
calculated independently from the samples described in Section~\ref{sec:mc}
and their uncertainties are combined in quadrature to obtain the \WZjj cross section prediction.
The predicted \EWWZ cross section is
$1.25^{+0.11}_{-0.09}\scale \pm 0.06\PDF \unit{fb}$,
and the interference term contribution in this region is less than 1\% of
the total cross section.

Results are also obtained in a looser fiducial region, defined in Table~\ref{tab:selections}
following Ref.~\cite{leshouches2017},
to simplify comparisons with theoretical calculations.
The acceptance from the loose to tight fiducial region
is $(72.4 \pm 0.8)\%$,
computed using \MG interfaced to \PYTHIA.
The uncertainty in the acceptance is evaluated
by combining the scale and PDF uncertainties
in the \EWWZ and \QCDWZ predictions in quadrature.
The scale uncertainty in the \QCDWZ contribution is the
dominant component of the uncertainty.
The resulting \WZjj loose fiducial cross section is
\begin{linenomath}
\begin{equation*}
  \sigma^{\mathrm{fid, loose}}_{\WZjj} =
        4.39^{+0.78}_{-0.72}\stat ^{+0.60}_{-0.50}\syst \unit{fb}
        = 4.39^{+0.98}_{-0.87} \unit{fb},
\end{equation*}
\end{linenomath}
compared with the predicted value of
$4.51^{+0.59}_{-0.45}\scale \pm 0.18\PDF \unit{fb}$.
The \EWWZ and \QCDWZ contributions
and their uncertainties are treated independently with the same approach as described
for the tight fiducial region.
The predicted \EWWZ cross section in the loose region is
$1.48^{+0.13}_{-0.11}\scale \pm 0.07\PDF \unit{fb}$,
and the relative contribution from the interference term is less the 1\%.

\begin{figure}[htbp]
  \centering
   \includegraphics[width=\cmsFigWidthSplit]{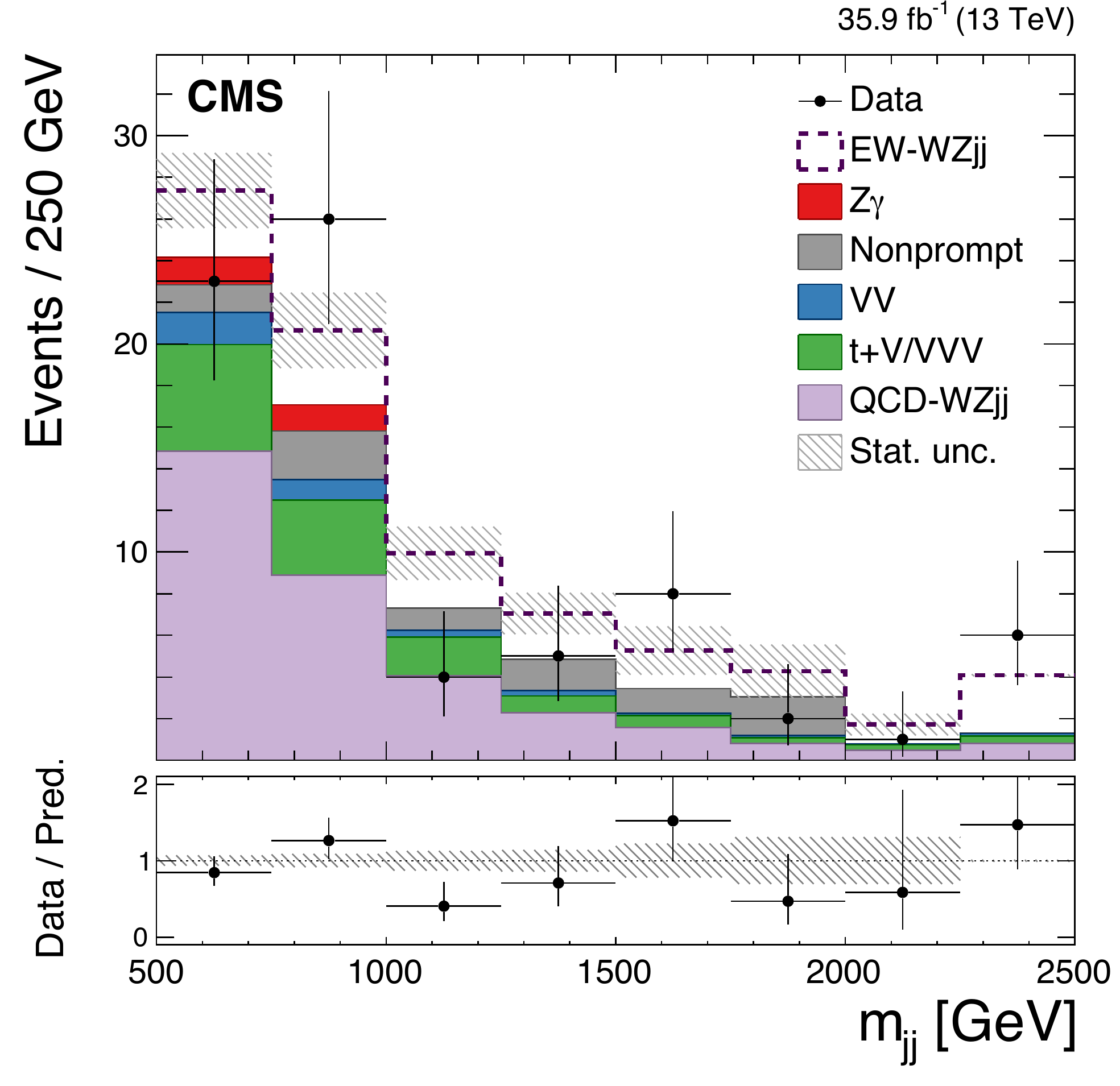}
   \includegraphics[width=\cmsFigWidthSplit]{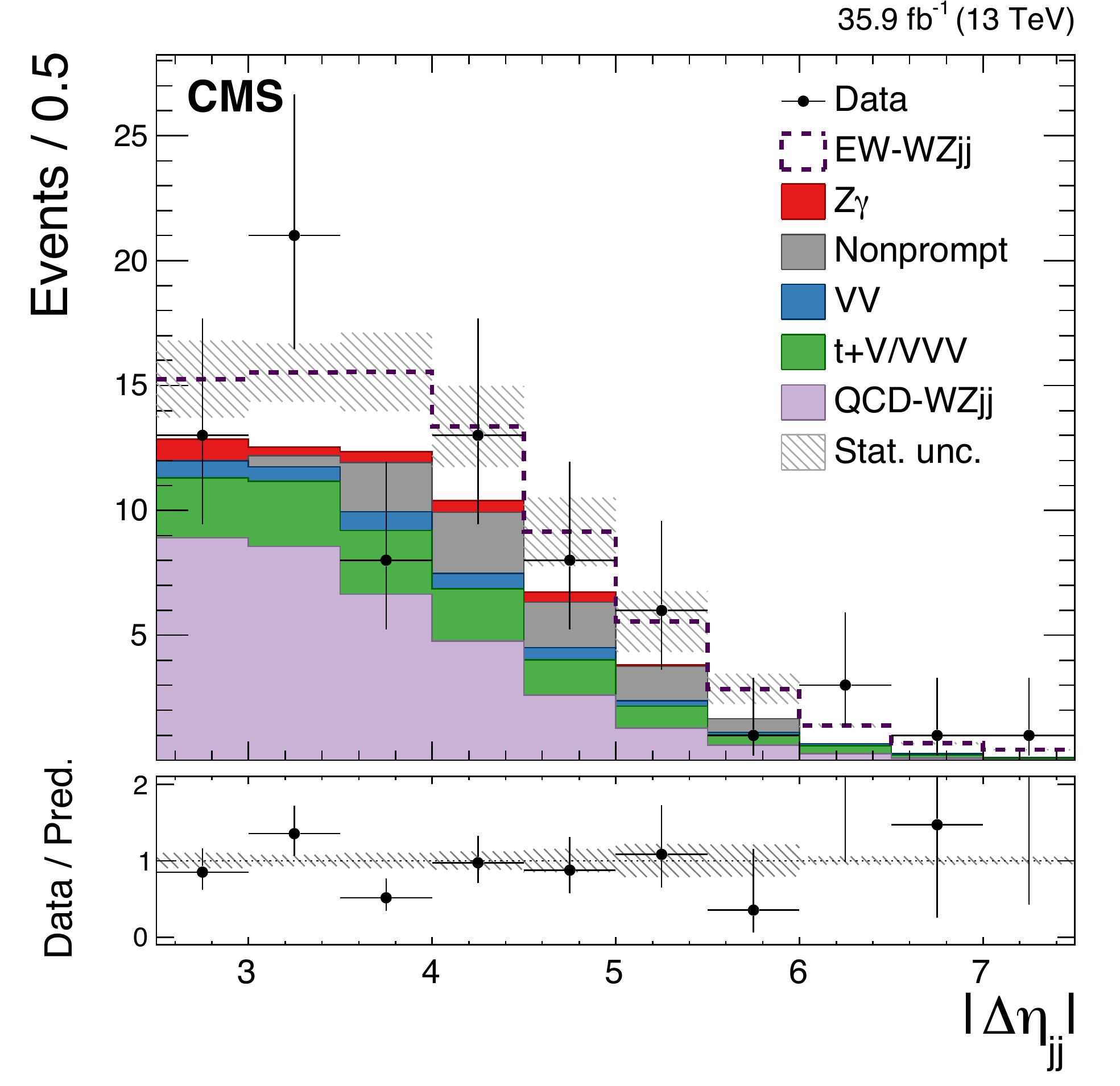}
  \caption{
    The $\mjj$ (\cmsLeft) and $\abs{\etajj}$ (\cmsRight)
  of the two leading jets
  for events satisfying the EW signal selection.
  The last bin contains all events with $\mjj > 2500\GeV$ (\cmsLeft) and
  $\abs{\etajj} > 7.5$ (\cmsRight).
  The dashed line shows the expected \EWWZ contribution stacked
  on top of the backgrounds that are shown as filled histograms.
  The hatched bands represent the total and relative
  statistical uncertainties on the predicted yields.
  The bottom panel shows the ratio of the number of events measured in data to the total
  number of expected events.
  The predicted yields are shown with their pre-fit normalizations.
          }
 \label{fig:VBSPlots}
\end{figure}

\begin{figure}[htbp]
  \centering
   \includegraphics[width=\cmsFigWidth]{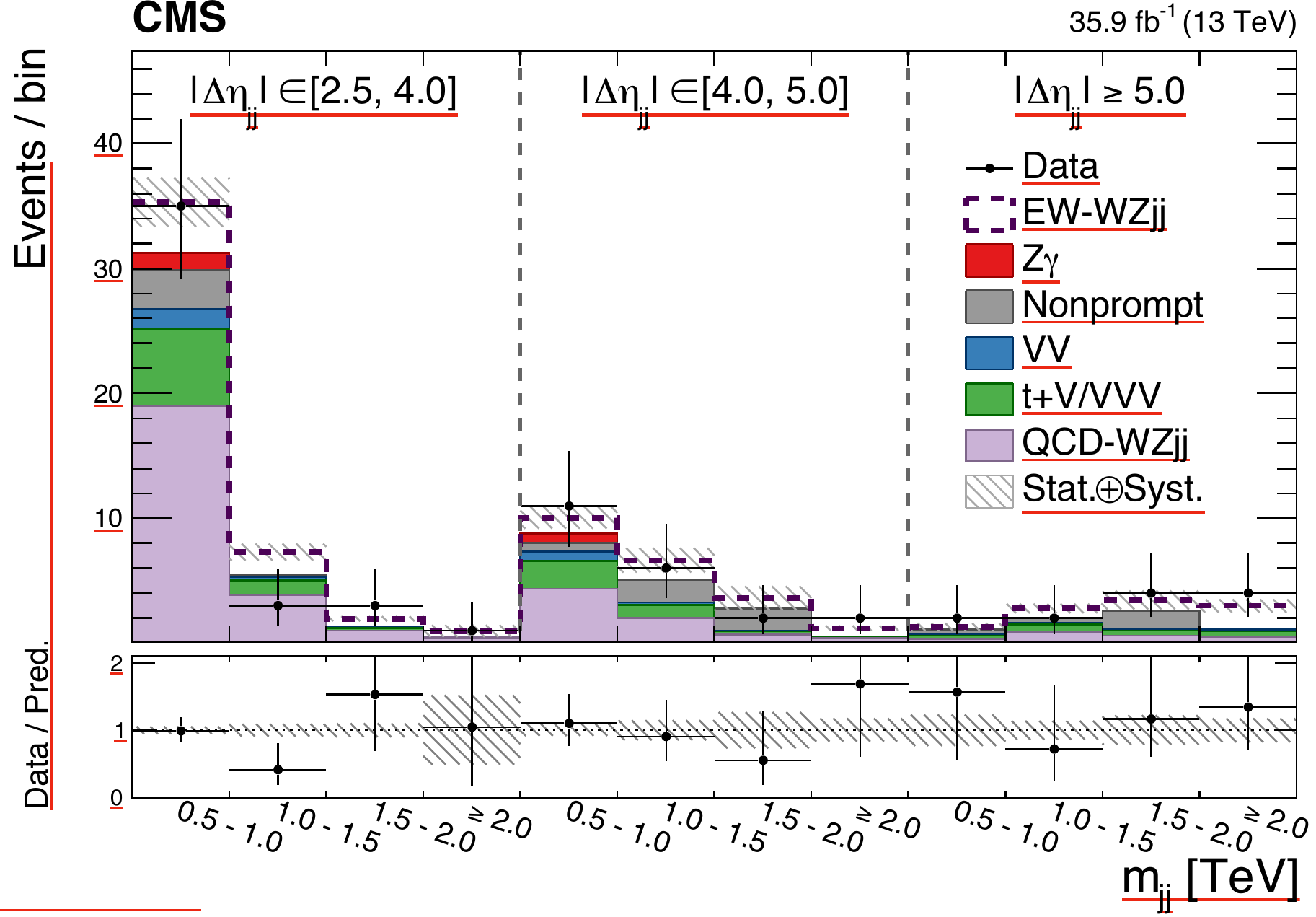}\qquad
  \caption{
    The one-dimensional representation of the 2D distribution of
    $\mjj$ and $\abs{\etajj}$, used for the EW
    signal extraction. The x axis shows the {\mjj} distribution
    in the indicated bins, split into three bins of {\etajj }: {\etajj} $\in [2.5, 4], [4, 5], \ge 5$.
    The dashed line represents the \EWWZ contribution stacked
    on top of the backgrounds that are shown as filled histograms.
    The hatched bands represent the total and relative
    systematic uncertainties on the predicted yields.
    The bottom panel shows the ratio of the number of events measured in data to the total
    number of expected events.
    The predicted yields are shown with their best-fit normalizations.
}
\label{fig:2DfitDistribution}
\end{figure}

Separating the EW- and QCD-induced components of \WZjj events requires exploiting the different
kinematic signatures of the two processes.
The relative fraction of the \EWWZ process with respect to the \QCDWZ process and other backgrounds
grows with increasing values of the $\mjj$ and $\abs{\etajj}$ of
the leading jets, as demonstrated in Fig.~\ref{fig:VBSPlots}.
This motivates the use of a 2D distribution built from these variables for the
extraction of the \EWWZ
signal via a maximum likelihood fit.
This 2D distribution,
shown as a one-dimensional histogram in Fig.~\ref{fig:2DfitDistribution},
along with the yield in the \QCDWZ sideband region, are combined in a binned likelihood
involving the expected and observed numbers of events in each bin.
The likelihood is a combination of individual likelihoods for the four decay channels.

The systematic uncertainties
are represented by nuisance parameters that are allowed to vary according to their
probability density functions, and correlation across bins and between different sources of uncertainty is taken
into account. The expected number of signal events is taken from the
\MG~v2.4.2 prediction at LO, multiplied by a signal strength $\mu_{\EW}$
which is treated as a
free parameter in the fit.

\begin{table*}[htbp]
  \centering
  \topcaption{Post-fit event yields after the signal extraction fit to events satisfying the EW signal selection.
    The \EWWZ process is corrected for the observed value of $\mu_{\EW}$.}
  \begin{tabular}{lccccc}
    \hline
        Process      &      \mmm      &      \emm      &      \eem      &      \eee     &  Total yield   \\
    \hline
        \QCDWZ       & 13.5 $\pm$ 0.8 & 9.1 $\pm$ 0.5  & 6.8 $\pm$ 0.4  & 4.6 $\pm$ 0.3 & 34.1 $\pm$ 1.1 \\
       t+V/\VVV      & 5.6 $\pm$ 0.4  & 3.1 $\pm$ 0.2  & 2.5 $\pm$ 0.2  & 1.7 $\pm$ 0.1 & 12.9 $\pm$ 0.5 \\
       Nonprompt     & 5.2 $\pm$ 2.0  & 2.4 $\pm$ 0.9  & 1.5 $\pm$ 0.6  & 0.7 $\pm$ 0.3 & 9.8 $\pm$ 2.3  \\
           VV        & 0.8 $\pm$ 0.1  & 1.6 $\pm$ 0.2  & 0.4 $\pm$ 0.0  & 0.7 $\pm$ 0.1 & 3.5 $\pm$ 0.2  \\
         $\Zg$       & 0.3 $\pm$ 0.1  & 1.2 $\pm$ 0.8  & $<$0.1         & 0.6 $\pm$ 0.2 & 2.2 $\pm$ 0.8  \\
      \hline
    Pred. background & 25.5 $\pm$ 2.1 & 17.4 $\pm$ 1.5 & 11.2 $\pm$ 0.8 & 8.3 $\pm$ 0.6 & 62.4 $\pm$ 2.8 \\
     \EWWZ signal    & 6.0 $\pm$ 1.2  & 4.2 $\pm$ 0.8  & 2.9 $\pm$ 0.6  & 2.1 $\pm$ 0.4 & 15.1 $\pm$ 1.6 \\
          Data       &       38       &       15       &       12       &       10      &       75       \\
    \hline
  \end{tabular}
  \label{tab:VBSYields}
\end{table*}

The best-fit value for the signal strength $\mu_{\EW}$ is
\begin{linenomath}
\begin{equation*}
  \mu_{\EW} = 0.82^{+0.51}_{-0.43},
\end{equation*}
\end{linenomath}
consistent with the SM expectation at LO of $\mu_{\EW,\,\mathrm{LO}} = 1$,
with respect to the predicted cross section for the \EWWZ process
in the tight fiducial region.
The significance of the signal is quantified by calculating the local $p$-value
for an upward fluctuation of the data relative to the background prediction
using a profile likelihood ratio test statistic and asymptotic formulae~\cite{Cowan:2010js}.
The observed\,(expected) statistical significance for \EWWZ production is 2.2\,(2.5) standard deviations.
A modification to the predicted cross section used in the fit trivially rescales
the signal strength but does not impact the significance of the result.
The total uncertainty of the
measurement is dominated by the statistical uncertainty of the data.
The post-fit yields for the signal and background corresponding to the best-fit signal strength
for \EWWZ production are shown in
Table~\ref{tab:VBSYields}.

\section{Limits on anomalous quartic gauge couplings}
\label{sec:aQGC}

Events satisfying the EW signal selection are used to constrain aQGCs in the effective field theory approach~\cite{Degrande:2012wf}.
Results are obtained following the formulation of Ref.~\cite{Eboli:2006wa} that proposes
nine independent dimension-eight operators, which
assume the {\SUtwo$\times$\Uone} symmetry of the EW gauge sector as well as
the presence of an SM Higgs boson. All operators are
charge conjugation and parity-conserving.
The \WZjj channel is most sensitive to the
T0, T1, and T2 operators that are constructed purely from the {\SUtwo} gauge fields,
the S0 and S1 operators that involve interactions with the Higgs field,
and the M0 and M1 operators that involve a mixture of gauge and Higgs field interactions.

\begin{figure}[htbp]
  \centering
    \includegraphics[width=\cmsFigWidth]{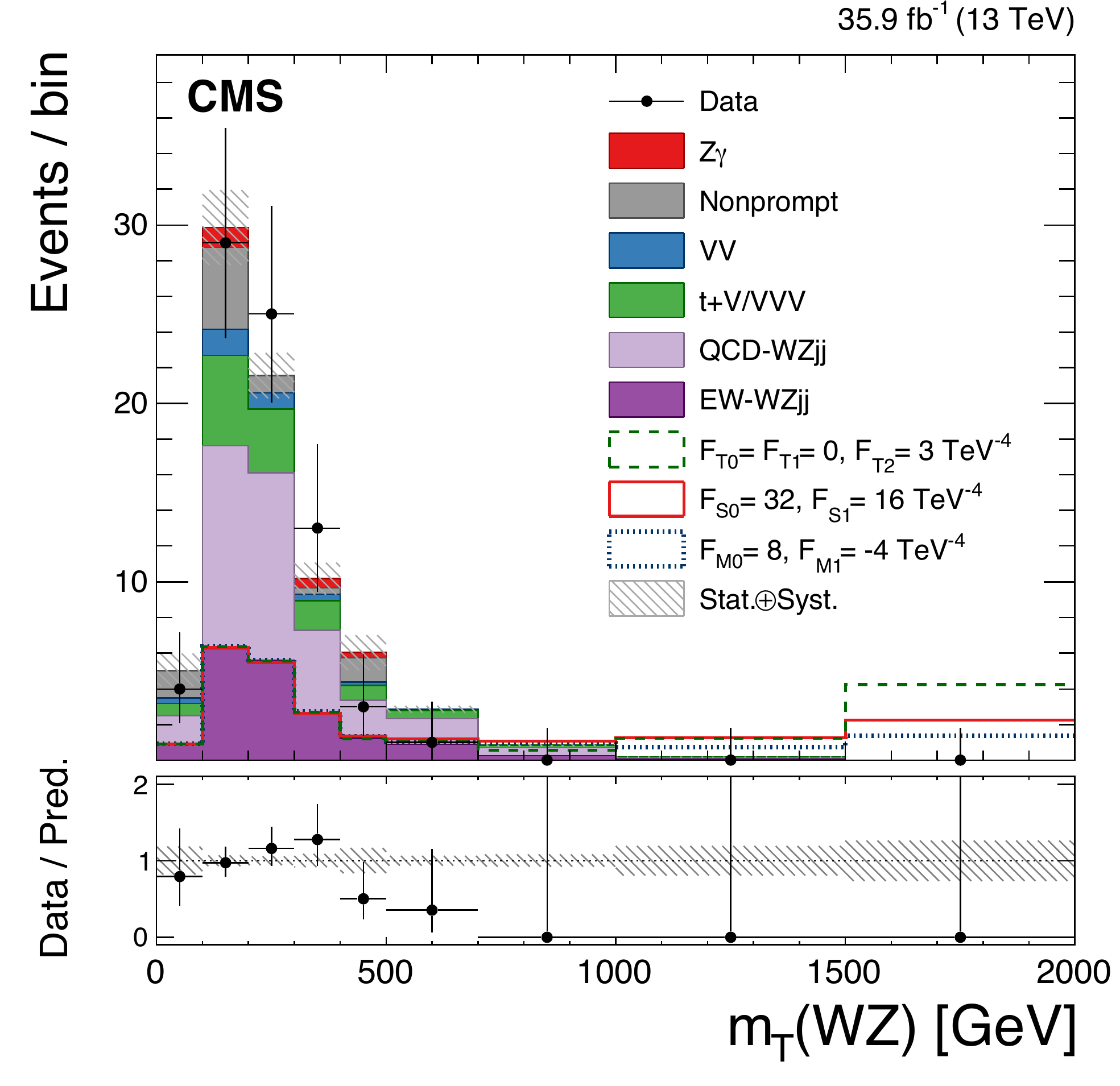}
  \caption{
      $\mt$ for events satisfying the EW signal selection,
      used to place constraints on the anomalous coupling parameters.
      The dashed lines show predictions for several aQGC parameters values that modify the \EWWZ process.
      The last bin contains all events with $\mt > 2000\GeV$.
      The hatched bands represent the total and relative
      systematic uncertainties on the predicted yields.
      The bottom panel shows the ratio of the number of events measured in data to the total
      number of expected events.
      The predicted yields are shown with their best-fit normalizations from the background-only fit.
      }
 \label{fig:aQGCDistribution}
\end{figure}

The presence of nonzero aQGCs would enhance the production of events with high
WZ mass. This motivates the use of the transverse mass of the \WZ system, defined as
\begin{linenomath}
\begin{equation*}
  \mt = \sqrt{\left[\ET(\PW) + \ET(\cPZ)\right]^2 - \left[\vec{\PT}(\PW) + \vec{\PT}(\cPZ)\right]^2},
  \label{eq:wztransmass}
\end{equation*}
\end{linenomath}
with $\ET = \sqrt{\smash[b]{m^2 + \PT^2}}$,
where the {\PW} candidate is constructed from
the \ptvecmiss
and the lepton associated with the $\PW$ boson,
and $m$ is the invariant mass of the $\PW$ or $\cPZ$ candidate,
to constrain the parameters $f_{\mathcal{O}i}/\Lambda^{4}$. In this formulation,
$f_{\mathcal{O}i}$ is a dimensionless coefficient for the operator $\mathcal{O}_{i}$ and $\Lambda$ is the energy scale of new physics.
The $\mt$ for events satisfying the
EW signal selection is shown in Fig~\ref{fig:aQGCDistribution}. The predictions of several
indicative aQGC operators and coefficients are also shown.

The MC simulations of nonzero aQGCs include the SM \EWWZ process, with an increase
in the yield at high $\mt$ arising from parameters different from their SM values.
Because the increase of the expected yield over the SM prediction exhibits a quadratic
dependence on the operator coefficient,
a parabolic function is fitted to the predicted yields per bin to obtain a smooth interpolation
between the discrete operator coefficients considered in the MC simulation.
The one-dimensional 95\% confidence level (\CL) limits are extracted
using the \CLs criterion~\cite{Junk:1999kv,CLS2,Cowan:2010js}, with all parameters
except for the coefficient being probed set to zero.
The SM prediction, including the \EWWZ process, is treated as the null hypothesis.
The expected prompt backgrounds are normalized to the predictions of
the MC simulations, with no corrections applied for the results of the \EWWZ or \WZjj measurements.
No deviation from the SM prediction is observed,
and the resulting observed and expected limits are summarized in Table~\ref{tab:1Dlimits}.

\begin{table} [htbp]
\centering
\topcaption{Observed and expected 95\% \CL limits for each operator
  coefficient (in TeV$^{-4}$) while all other parameters are set to zero.}
\begin{tabular}{ccc}
\hline
  Parameters & Exp. limit & Obs. limit  \\
\hline
f$_{\text{M0}}/\Lambda^4$ & $[-11.2, 11.6]$ & $[-9.15, 9.15]$ \\
f$_{\text{M1}}/\Lambda^4$ & $[-10.9, 11.6]$ & $[-9.15, 9.45]$ \\
f$_{\text{S0}}/\Lambda^4$ & $[-32.5, 34.5]$ & $[-26.5, 27.5]$ \\
f$_{\text{S1}}/\Lambda^4$ & $[-50.2, 53.2]$ & $[-41.2, 42.8]$ \\
f$_{\text{T0}}/\Lambda^4$ & $[-0.87, 0.89]$ & $[-0.75, 0.81]$ \\
f$_{\text{T1}}/\Lambda^4$ & $[-0.56, 0.60]$ & $[-0.49, 0.55]$ \\
f$_{\text{T2}}/\Lambda^4$ & $[-1.78, 2.00]$ & $[-1.49, 1.85]$ \\
\hline
\end{tabular}
\label{tab:1Dlimits}
\end{table}

\begin{figure}[htb!]
  \centering
   \includegraphics[width=\cmsFigWidthSplit]{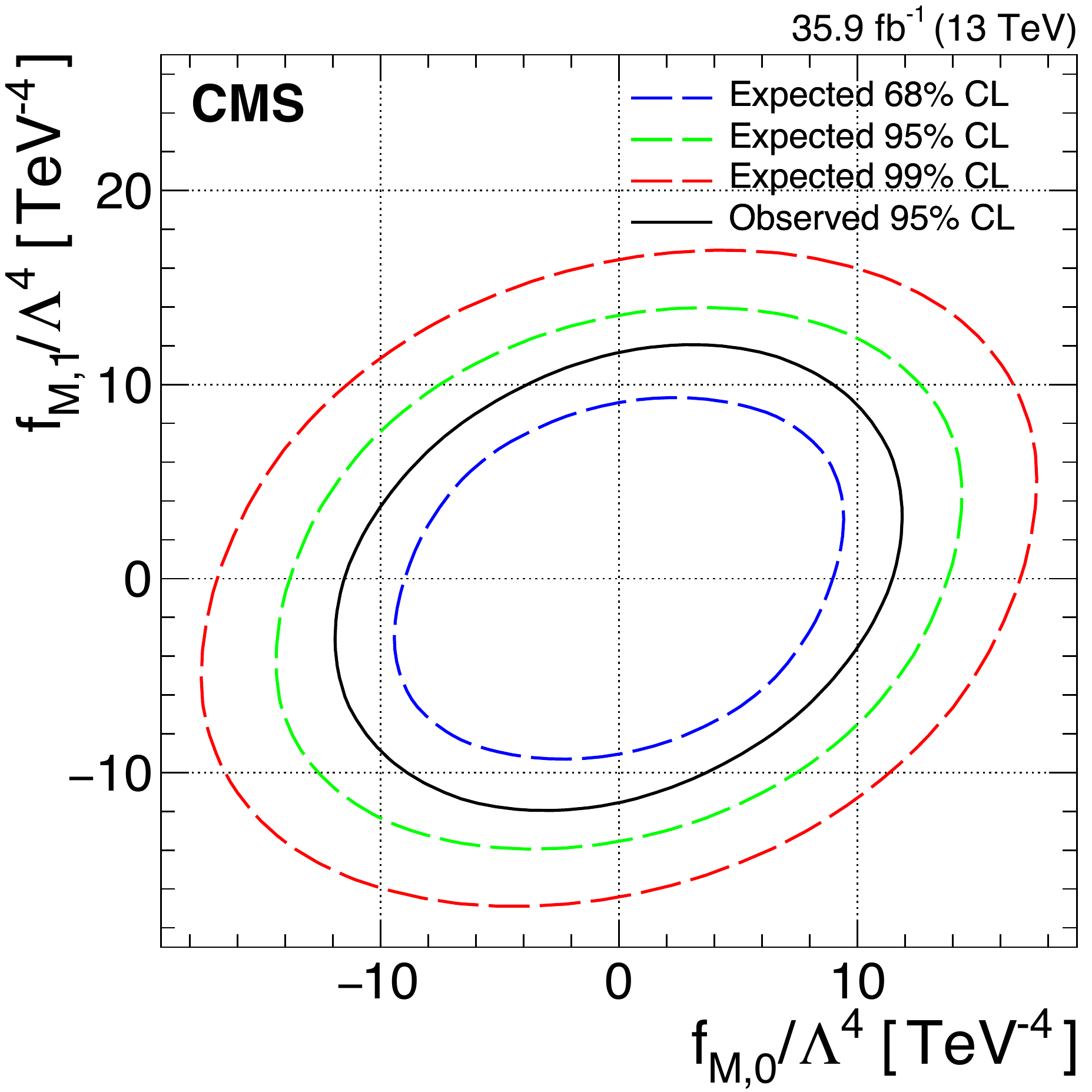}
   \includegraphics[width=\cmsFigWidthSplit]{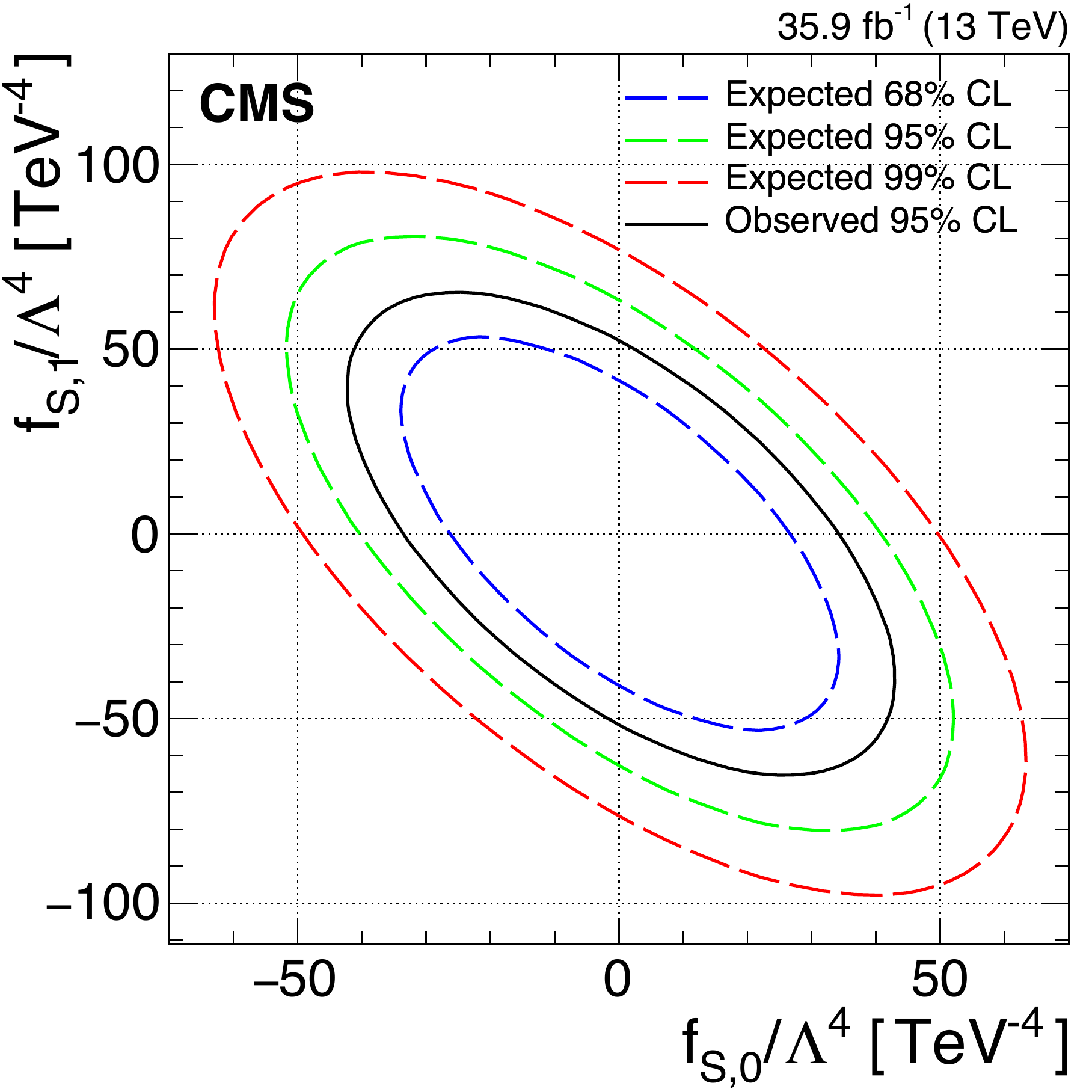}
\caption{Two-dimensional observed 95\% \CL intervals (solid contour) and expected
68, 95, and 99\% \CL intervals (dashed contour) on the selected aQGC parameters.
The values of coefficients
outside of contours are excluded at the corresponding \CL.
 }
 \label{fig:2Dlimits}
\end{figure}

Constraints are also placed on aQGC parameters using a two-dimensional scan,
where two parameters are probed in the fit with all others set to zero.
This approach is motivated by correlations between operators and physical couplings,
and for comparisons with alternative formulations of dimension-eight operators.
In particular, the quartic gauge interactions of the massive gauge bosons
is a function of S0 and S1, while combinations of the M0 and M1 operators can be
compared with the formulation of Ref.~\cite{Belanger:1992qh}.
The resulting 2D 95\% \CL intervals for these parameters are shown in Fig.~\ref{fig:2Dlimits}.

\section{Limits on charged Higgs boson production}
\label{sec:chargedHiggs}

Theories with Higgs sectors including {\SUtwo} triplets can give rise to charged Higgs
bosons (H$^{\pm}$) with large couplings to the vector bosons of the SM.
A prominent one is the GM model~\cite{GEORGI1985463},
where the Higgs sector is extended by one real and one complex {\SUtwo} triplet to preserve custodial
symmetry at tree level for arbitrary vacuum expectation values.
In this model, the couplings of $\PH^{\pm}$ and the vector bosons depend on
$m(\PHpm)$ and the parameter $\sin{\theta_{\PH}}$, or $s_{\PH}$,
which represents the mixing angle of the vacuum expectation values in the model, and
determines the fraction of the {\PW} and {\cPZ} boson masses generated by the
vacuum expectation values of the triplets.
This analysis extends the previous study of $\PH^{\pm}$ production via
vector boson fusion
by the CMS Collaboration in the same channel~\cite{Sirunyan:2017sbn}.

\begin{figure}[htb]
  \centering
    \includegraphics[width=\cmsFigWidth]{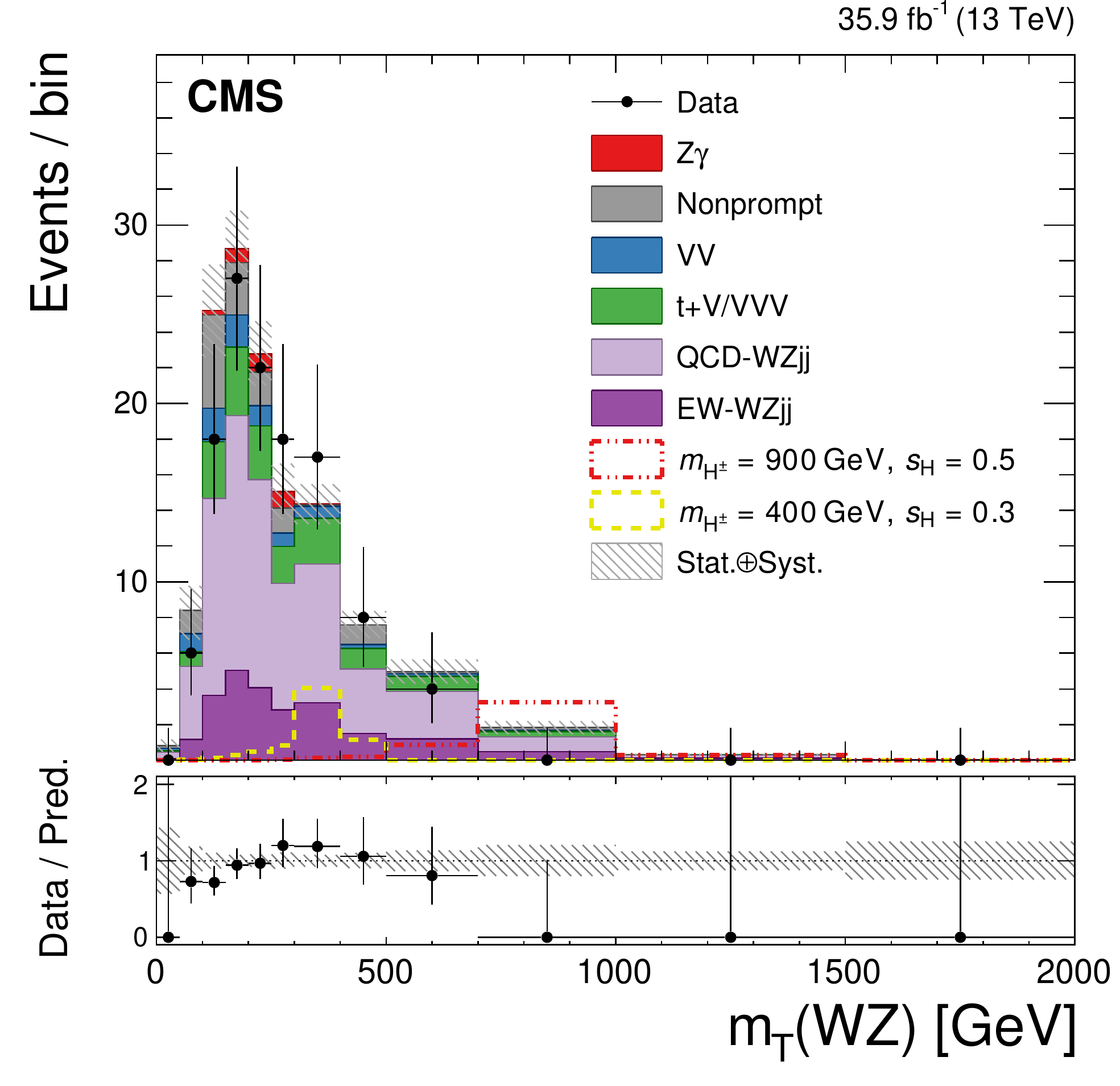}
  \caption{
      $\mt$ for events satisfying the Higgs boson selection,
      used to place constraints on the production of charged Higgs bosons.
      The last bin contains all events with $\mt > 2000\GeV$.
      The dashed lines show predictions from the GM model with
      $m(\PHpm) = 400 \,(900)\GeV$ and $s_{\PH} = 0.3 \,(0.5)$.
      The bottom panel shows the ratio of the number of events measured in data to the total
      number of expected events. The hatched bands represent the total and relative
      systematic uncertainties on the predicted background yields.
      The predicted yields are shown with their best-fit normalizations from the background-only fit.
      }
 \label{fig:higgsmt}
\end{figure}

\begin{figure}[htb]
  \centering
  \includegraphics[width=\cmsFigWidthSplit]{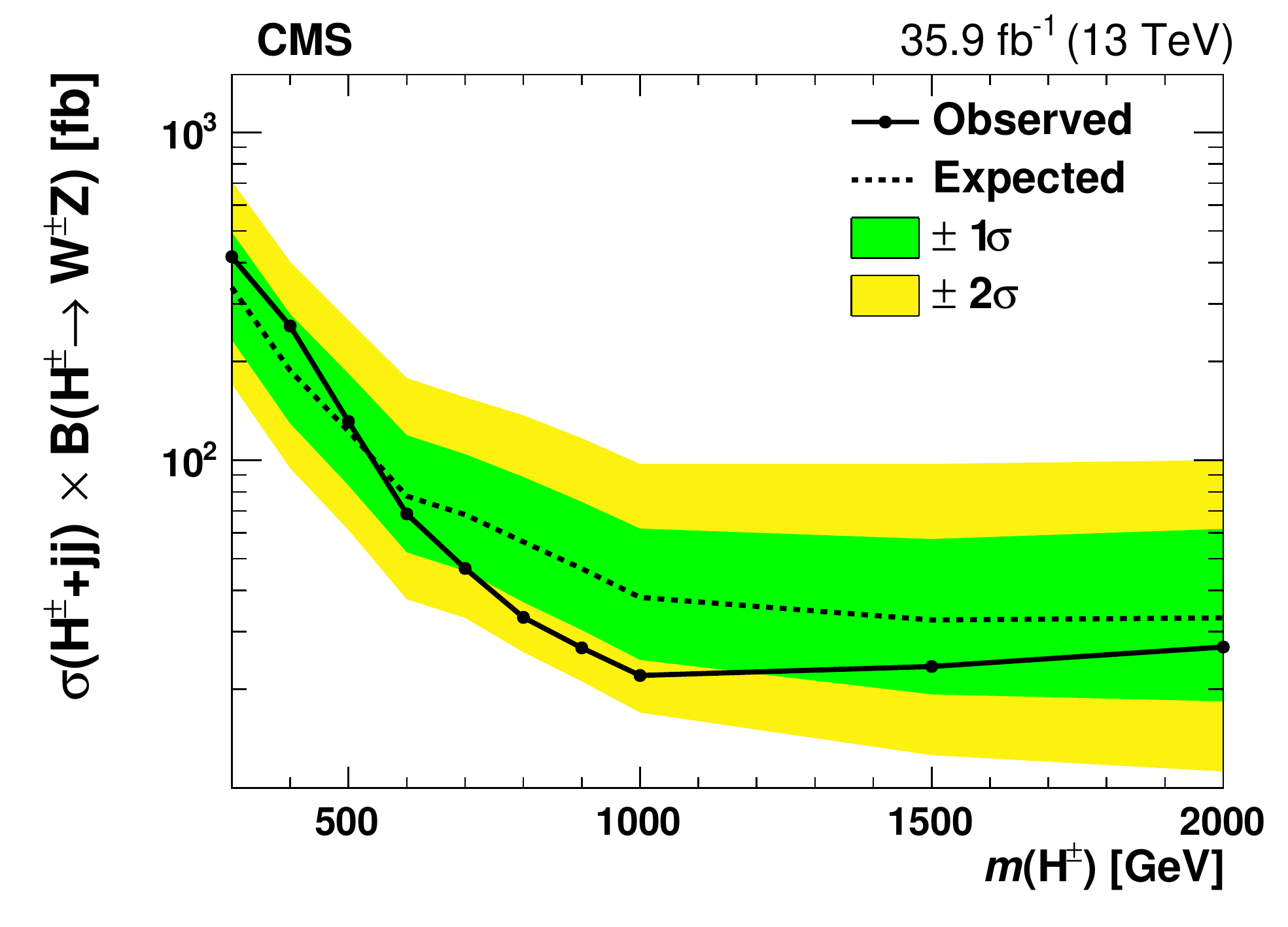}
  \includegraphics[width=\cmsFigWidthSplit]{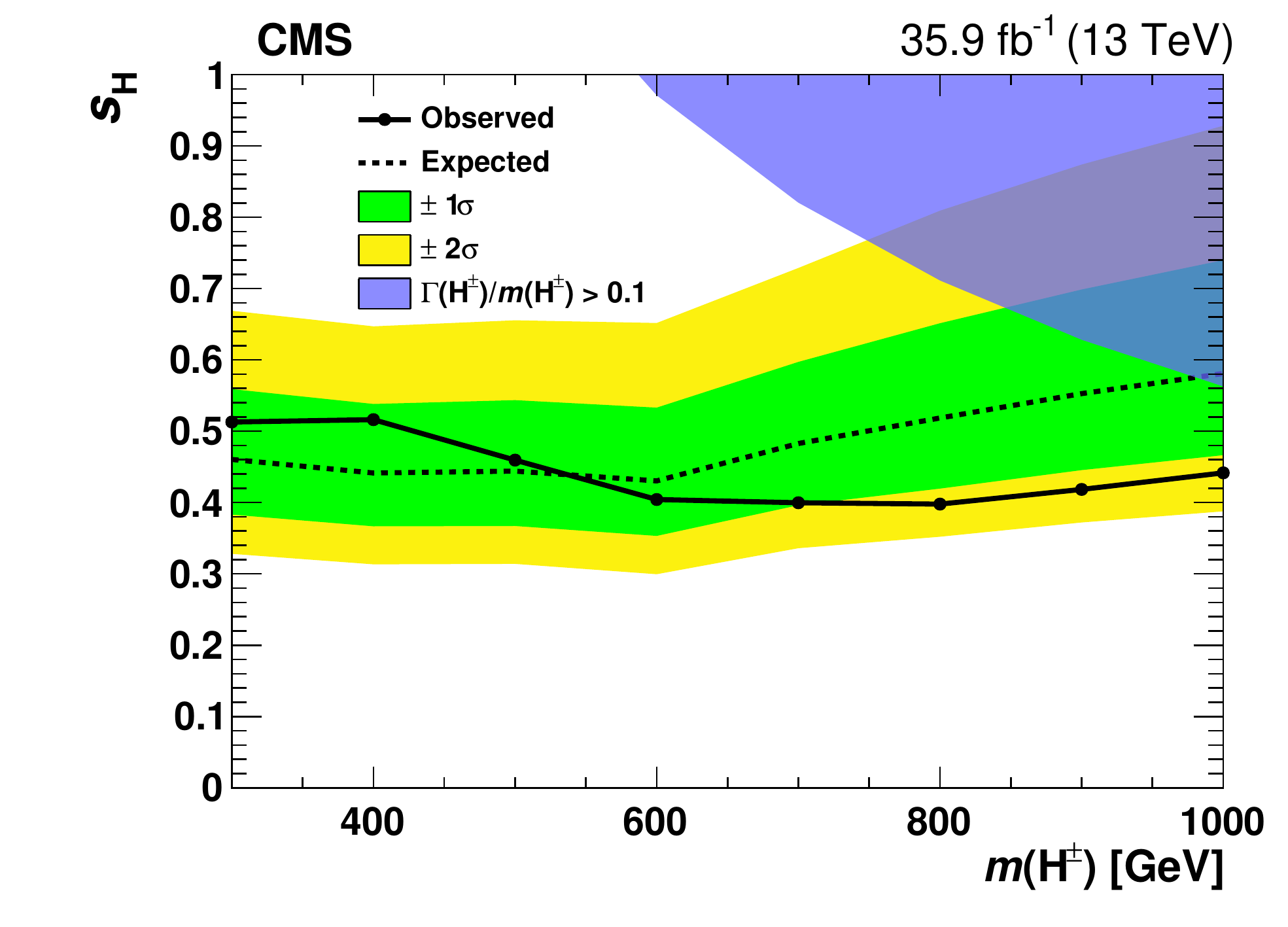}
\caption{Expected (dashed lines)
  and observed (solid lines) upper limits at 95\% \CL for the model independent
  $\sigma(\Hpjj) \, \mathcal{B}(\PHpm\to \PW\cPZ)$
  as a function of $m(\PH^\pm)$ (\cmsLeft) and
  for $s_{\PH}$ as a function of $m_{\PH}$ in the GM model (\cmsRight).
  The blue shaded area covers the theoretically not allowed parameter space~\cite{Zaro:2002500}.
}
\label{fig:limits}
\end{figure}

A combined fit of the predicted signal and background yields to the data
in the Higgs boson selection
is performed in bins of $\mt$, simultaneously with the event yield in the \QCDWZ sideband region,
to derive model-independent expected and observed upper limits on 
$\sigma(\Hpjj) \,\mathcal{B}(\PHpm\to \PW\cPZ)$ at 95\% \CL using the \CLs criterion.
The distribution and binning of the $\mt$ distribution used in the fit are shown in Fig.~\ref{fig:higgsmt}.
The upper limits as a function of $m(\PHpm)$
are shown in Fig.~\ref{fig:limits} (\cmsLeft).
The results assume that the intrinsic width of the $\PHpm$ is $\lesssim$0.05$m(\PHpm)$,
which is below the experimental resolution in the phase space considered.

The model-independent upper limits are compared with the predicted cross sections at next-to-next-to-leading order in the
GM model in the $s_{\PH}$-$m(\PHpm)$ plane,
under the assumptions defined for the ``H5plane'' in Ref.~\cite{Zaro:2002500}. For the probed parameter space and
$\mt$ distribution used for signal extraction, the varying width as a function of $s_{\PH}$ is assumed
to have negligible effect on the result. The value of the branching fraction $\mathcal{B}(\PHpm\to \PW\cPZ)$ is
assumed to be unity. In Fig.~\ref{fig:limits} (\cmsRight), the excluded $s_{\PH}$ values as a function of
$m(\PHpm)$ are shown. The blue shaded region shows the parameter space for which the $\PHpm$ total width
exceeds 10\% of $m(\PHpm)$, where the model is not applicable because of perturbativity and vacuum
stability requirements~\cite{Zaro:2002500}.

\section{Summary}
\label{sec:summary}
A measurement of the production of a {\PW} and a $\PZ$ boson in association with two jets has been presented,
using events where both bosons decay leptonically.
Results are based on data corresponding to an integrated luminosity of $35.9\fbinv$
recorded in proton-proton collisions at $\sqrt{s} = 13\TeV$ with the CMS detector
at the LHC in 2016. The cross section in a tight fiducial region with enhanced contributions from
electroweak (EW) \WZ production is $\sigma^{\mathrm{fid}}_{\WZjj} = 3.18^{+0.71}_{-0.63}\unit{fb}$,
consistent with the standard model (SM) prediction.
The dijet mass and dijet rapidity separation are used to measure
the signal strength of \EWWZ production with
respect to the SM expectation, resulting in
$\mu_{\EW} = 0.82^{+0.51}_{-0.43}$.
The significance of this result is
2.2 standard deviations with 2.5 standard deviations expected.

Constraints are placed on anomalous quartic gauge couplings
in terms of dimension-eight effective field theory operators, and
upper limits are given on the production cross section
times branching fraction of charged Higgs bosons.
The upper limits on charged Higgs boson production
via vector boson fusion with decay to a {\PW} and a {\cPZ} boson
extend the results previously published
by the CMS Collaboration~\cite{Sirunyan:2017sbn} and
are comparable to those of the ATLAS Collaboration~\cite{Aaboud:2018ohp}.
These are the first limits for dimension-eight effective field theory
operators in the \WZ channel at $13\TeV$.

\begin{acknowledgments}

We congratulate our colleagues in the CERN accelerator departments for the excellent performance of the LHC and thank the technical and administrative staffs at CERN and at other CMS institutes for their contributions to the success of the CMS effort. In addition, we gratefully acknowledge the computing centers and personnel of the Worldwide LHC Computing Grid for delivering so effectively the computing infrastructure essential to our analyses. Finally, we acknowledge the enduring support for the construction and operation of the LHC and the CMS detector provided by the following funding agencies: BMBWF and FWF (Austria); FNRS and FWO (Belgium); CNPq, CAPES, FAPERJ, FAPERGS, and FAPESP (Brazil); MES (Bulgaria); CERN; CAS, MoST, and NSFC (China); COLCIENCIAS (Colombia); MSES and CSF (Croatia); RPF (Cyprus); SENESCYT (Ecuador); MoER, ERC IUT, and ERDF (Estonia); Academy of Finland, MEC, and HIP (Finland); CEA and CNRS/IN2P3 (France); BMBF, DFG, and HGF (Germany); GSRT (Greece); NKFIA (Hungary); DAE and DST (India); IPM (Iran); SFI (Ireland); INFN (Italy); MSIP and NRF (Republic of Korea); MES (Latvia); LAS (Lithuania); MOE and UM (Malaysia); BUAP, CINVESTAV, CONACYT, LNS, SEP, and UASLP-FAI (Mexico); MOS (Montenegro); MBIE (New Zealand); PAEC (Pakistan); MSHE and NSC (Poland); FCT (Portugal); JINR (Dubna); MON, RosAtom, RAS, RFBR, and NRC KI (Russia); MESTD (Serbia); SEIDI, CPAN, PCTI, and FEDER (Spain); MOSTR (Sri Lanka); Swiss Funding Agencies (Switzerland); MST (Taipei); ThEPCenter, IPST, STAR, and NSTDA (Thailand); TUBITAK and TAEK (Turkey); NASU and SFFR (Ukraine); STFC (United Kingdom); DOE and NSF (USA).

\hyphenation{Rachada-pisek} Individuals have received support from the Marie-Curie program and the European Research Council and Horizon 2020 Grant, contract No. 675440 (European Union); the Leventis Foundation; the A.P.\ Sloan Foundation; the Alexander von Humboldt Foundation; the Belgian Federal Science Policy Office; the Fonds pour la Formation \`a la Recherche dans l'Industrie et dans l'Agriculture (FRIA-Belgium); the Agentschap voor Innovatie door Wetenschap en Technologie (IWT-Belgium); the F.R.S.-FNRS and FWO (Belgium) under the ``Excellence of Science -- EOS" -- be.h project n.\ 30820817; the Ministry of Education, Youth and Sports (MEYS) of the Czech Republic; the Lend\"ulet (``Momentum") Programme and the J\'anos Bolyai Research Scholarship of the Hungarian Academy of Sciences, the New National Excellence Program \'UNKP, the NKFIA research grants 123842, 123959, 124845, 124850, and 125105 (Hungary); the Council of Science and Industrial Research, India; the HOMING PLUS programme of the Foundation for Polish Science, cofinanced from European Union, Regional Development Fund, the Mobility Plus programme of the Ministry of Science and Higher Education, the National Science Center (Poland), contracts Harmonia 2014/14/M/ST2/00428, Opus 2014/13/B/ST2/02543, 2014/15/B/ST2/03998, and 2015/19/B/ST2/02861, Sonata-bis 2012/07/E/ST2/01406; the National Priorities Research Program by Qatar National Research Fund; the Programa Estatal de Fomento de la Investigaci{\'o}n Cient{\'i}fica y T{\'e}cnica de Excelencia Mar\'{\i}a de Maeztu, grant MDM-2015-0509 and the Programa Severo Ochoa del Principado de Asturias; the Thalis and Aristeia programmes cofinanced by EU-ESF and the Greek NSRF; the Rachadapisek Sompot Fund for Postdoctoral Fellowship, Chulalongkorn University and the Chulalongkorn Academic into Its 2nd Century Project Advancement Project (Thailand); the Welch Foundation, contract C-1845; and the Weston Havens Foundation (USA).

\end{acknowledgments}

\bibliography{auto_generated}
\cleardoublepage \appendix\section{The CMS Collaboration \label{app:collab}}\begin{sloppypar}\hyphenpenalty=5000\widowpenalty=500\clubpenalty=5000\input{SMP-18-001-authorlist.tex}\end{sloppypar}
\end{document}

%% file: SMP-18-001-authorlist.tex
\vskip\cmsinstskip
\textbf{Yerevan Physics Institute, Yerevan, Armenia}\\*[0pt]
A.M.~Sirunyan, A.~Tumasyan
\vskip\cmsinstskip
\textbf{Institut f\"{u}r Hochenergiephysik, Wien, Austria}\\*[0pt]
W.~Adam, F.~Ambrogi, E.~Asilar, T.~Bergauer, J.~Brandstetter, M.~Dragicevic, J.~Er\"{o}, A.~Escalante~Del~Valle, M.~Flechl, R.~Fr\"{u}hwirth\cmsAuthorMark{1}, V.M.~Ghete, J.~Hrubec, M.~Jeitler\cmsAuthorMark{1}, N.~Krammer, I.~Kr\"{a}tschmer, D.~Liko, T.~Madlener, I.~Mikulec, N.~Rad, H.~Rohringer, J.~Schieck\cmsAuthorMark{1}, R.~Sch\"{o}fbeck, M.~Spanring, D.~Spitzbart, W.~Waltenberger, J.~Wittmann, C.-E.~Wulz\cmsAuthorMark{1}, M.~Zarucki
% ...
\vskip\cmsinstskip
% ...
1:  Also at Vienna University of Technology, Vienna, Austria\\
% ...
